\documentclass[aip,amsmath,amssymb,preprint,]{revtex4-1}

\usepackage{graphicx}
\usepackage{subcaption}
\usepackage{dcolumn}
\usepackage{bm}

\usepackage[utf8]{inputenc}
\usepackage[T1]{fontenc}
\usepackage{mathptmx}
\usepackage{mathrsfs}
\usepackage{etoolbox}
\usepackage{amsthm,amsmath,amssymb}
\usepackage{natbib}
\usepackage{booktabs}
\usepackage{CJKutf8}
\DeclareMathAlphabet{\mathcal}{OMS}{cmsy}{m}{n}
\DeclareSymbolFont{largesymbols}{OMX}{cmex}{m}{n}
 
\begin{document}
\preprint{AIP/123-QED}

\title[]{Prediction of three-dimensional chemically reacting compressible turbulence based on implicit U-Net
 enhanced Fourier neural operator}
\author{Zhiyao Zhang(\begin{CJK}{UTF8}{gbsn}张志尧\end{CJK})}
\affiliation{Department of Mechanics and Aerospace Engineering, 
Southern University of Science and Technology, Shenzhen 518055, 
China}
\affiliation{Guangdong-Hong Kong-Macao Joint Laboratory for Data-Driven Fluid Mechanics 
and Engineering Applications, Southern University of Science and Technology, 
Shenzhen 518055, China}
\affiliation{Guangdong Provincial Key Laboratory of Turbulence Research and 
Applications, Department of Mechanics and Aerospace Engineering, 
Southern University of Science and Technology, Shenzhen 518055, 
China}
\author{Zhijie Li(\begin{CJK}{UTF8}{gbsn}李志杰\end{CJK})}
\affiliation{Department of Mechanics and Aerospace Engineering, 
Southern University of Science and Technology, Shenzhen 518055, 
China}
\affiliation{Guangdong-Hong Kong-Macao Joint Laboratory for Data-Driven Fluid Mechanics 
and Engineering Applications, Southern University of Science and Technology, 
Shenzhen 518055, China}
\affiliation{Guangdong Provincial Key Laboratory of Turbulence Research and 
Applications, Department of Mechanics and Aerospace Engineering, 
Southern University of Science and Technology, Shenzhen 518055, 
China}

\author{Yunpeng Wang(\begin{CJK}{UTF8}{gbsn}王云朋\end{CJK})}
\affiliation{Department of Mechanics and Aerospace Engineering, 
Southern University of Science and Technology, Shenzhen 518055, 
China}
\affiliation{Guangdong-Hong Kong-Macao Joint Laboratory for Data-Driven Fluid Mechanics 
and Engineering Applications, Southern University of Science and Technology, 
Shenzhen 518055, China}
\affiliation{Guangdong Provincial Key Laboratory of Turbulence Research and 
Applications, Department of Mechanics and Aerospace Engineering, 
Southern University of Science and Technology, Shenzhen 518055, 
China}

\author{Huiyu Yang(\begin{CJK}{UTF8}{gbsn}阳汇昱\end{CJK})}
\affiliation{Department of Mechanics and Aerospace Engineering, 
Southern University of Science and Technology, Shenzhen 518055, 
China}
\affiliation{Guangdong-Hong Kong-Macao Joint Laboratory for Data-Driven Fluid Mechanics 
and Engineering Applications, Southern University of Science and Technology, 
Shenzhen 518055, China}
\affiliation{Guangdong Provincial Key Laboratory of Turbulence Research and 
Applications, Department of Mechanics and Aerospace Engineering, 
Southern University of Science and Technology, Shenzhen 518055, 
China}

\author{Wenhui Peng(\begin{CJK}{UTF8}{gbsn}彭文辉\end{CJK})}
\affiliation{Hubei University of Medicine, Hubei 442000, China}

\author{Jian Teng(\begin{CJK}{UTF8}{gbsn}滕健\end{CJK})}
\affiliation{School of Ocean Engineering, 
Guangzhou Maritime University, 
Guangzhou 510725, Guangdong, People's Republic of China.
}

\author{Jianchun Wang(\begin{CJK}{UTF8}{gbsn}王建春\end{CJK})}
\email{wangjc@sustech.edu.cn}
\affiliation{Department of Mechanics and Aerospace Engineering, 
 Southern University of Science and Technology, Shenzhen 518055, 
 China}
\affiliation{Guangdong-Hong Kong-Macao Joint Laboratory for Data-Driven Fluid Mechanics 
 and Engineering Applications, Southern University of Science and Technology, 
 Shenzhen 518055, China}
\affiliation{Guangdong Provincial Key Laboratory of Turbulence Research and 
 Applications, Department of Mechanics and Aerospace Engineering, 
 Southern University of Science and Technology, Shenzhen 518055, 
 China}

\date{\today}

\begin{abstract}
  The accurate and fast prediction of long-term dynamics of turbulence presents a significant challenge for 
  both traditional numerical simulations and machine learning methods. In recent years, the 
  emergence of neural operators 
  has provided a promising approach to address this issue. The implicit U-Net enhanced 
  Fourier neural operator (IU-FNO) has 
  successfully demonstrated long-term stable predictions for three-dimensional incompressible 
  turbulence. In this study, we extend this method to the 
  three-dimensional chemically reacting compressible turbulence. Numerical results show that 
  the IU-FNO model predicts flow dynamics significantly faster than the traditional dynamic 
  Smagorinsky model (DSM) 
  used in large eddy simulation (LES). In terms of prediction accuracy, the IU-FNO 
  framework outperforms the traditional DSM in predicting the energy spectra of velocity, 
  temperature, and density, the probability density functions (PDFs) of vorticity and 
  velocity increments, and instantaneous spatial structures of temperature. Therefore, the IU-FNO represents a 
  highly promising approach for predicting chemically reacting compressible turbulence.
  
\end{abstract}

\maketitle

\begin{quotation}
\end{quotation}

\section{introduction}
Turbulent flows are ubiquitous in natural phenomena and anthropogenic activities, 
including meteorology, air pollution, aerospace engineering, and industrial production.\cite{pope2000turbulent}
There are three primary numerical methods for turbulence: 
direct numerical simulation (DNS)\cite{kaneda2006high,moin1998direct}, Reynolds-averaged Navier-Stokes (RANS)\cite{alfonsi2009reynolds,chen1990solutions} simulations, 
and large eddy simulation (LES)\cite{sagaut2005large,lesieur1996new}.
Since the DNS method resolves all scales of the flow field, its application at high Reynolds 
numbers remains impractical.\cite{meneveau2000scale,moser2021statistical}
RANS methods employ time-averaging to eliminate the need for resolving turbulent fluctuations and significantly reduce computational cost, hence they are widely applied
in industrial applications.
\cite{duraisamy2019turbulence,jiang2021interpretable,durbin2018some,pope1975more,chen2023universal,bin2024constrained}
LES adopts a compromise approach by directly resolving large-scale flow 
structures while modeling the smaller scales, thus reducing computational cost while preserving 
the fluctuating nature of the turbulent flow.
\cite{smagorinsky1963general,lilly1967representation,deardorff1970numerical,sagaut2005large,moin1991dynamic,germano1992turbulence,mons2021ensemble,rouhi2016dynamic,celora2021covariant,rozema2022local}
Numerous subgrid-scale (SGS) models have been proposed for large eddy simulation, including 
the Smagorinsky model(SM)\cite{smagorinsky1963general}, the dynamic Smagorinsky model (DSM)\cite{smagorinsky1963general}, the velocity gradient model\cite{vreman1996large}, 
the scale-similarity models\cite{stallcup2022adaptive}, the dynamic mixed model (DMM)\cite{vreman1994formulation,vreman1997large}, 
and the dynamic SGS models based on multiscale properties of turbulence\cite{yuan2021dynamic,wang2021artificial}.

Recently, various machine learning-based methods for flow field prediction have emerged.\cite{maulik2018data,wang2018investigations,yang2019predictive,zhou2019subgrid,beck2019deep,brunton2020machine,park2021toward,karniadakis2021physics,xu2023practical,guan2022stable,bae2022scientific,wu2024spatio,li2024synthetic,xu2022pde,xiao2020flows,kurz2020machine,altland2022modeling,duraisamy2021perspectives,deng2023temporal,lozano2023machine,lienen2023zero,li2024scalable,wu2024transolver,gao2024generative}
Particularly, increasing attention is being 
focused on machine learning approach for direct predictions of the entire flow field. The relevant approaches include recurrent neural networks and long short-term memory methods
\cite{bukka2021assessment,han2022deep},
physics-informed neural network (PINN) methods\cite{raissi2019physics,jin2021nsfnets}, 
and neural operators\cite{li2020fourier,lu2019deeponet,lu2021learning,wang2021learning,deng2023temporal,peng2022attention,peng2023linear,luo2024fourier,hao2023instability,ranade2021generalized}.
For example, Han et al. utilized convolutional neural networks and long short-term memory 
frameworks to predict fluid-structure interaction\cite{han2021hybrid,han2022deep}. 
Raissi et al. introduced the concept of incorporating physical equations to constrain 
the loss function, achieving promising results in the solution of various nonlinear partial 
differential equations\cite{raissi2019physics}.
Wang et al. introduced trainable spectral filters into the coupled model of Reynolds-averaged 
Navier-Stokes (RANS) and large eddy simulation (LES), and subsequently employed a corresponding 
U-Net architecture to predict meteorological data\cite{wang2020towards}.

However, most neural networks establish input-output pairs directly in finite-dimensional 
Euclidean space, which makes it challenging for these methods to generalize to more complex 
flows and varying boundary conditions. The Fourier neural operator (FNO) proposed by Li et al.\cite{li2020fourier} allows 
for learning the mapping of datasets in infinite space by transforming the data into Fourier space.
The FNO has achieved an accurate 
predictive performance in various scenarios, including the Burgers equation, 
Darcy flow, and the Navier-Stokes equations. Since Li et al., the FNO method has been widely applied 
in turbulence-related studies\cite{wen2022u,peng2023linear,you2022learning,zhang2024learning,tran2021factorized,qin2024toward,cao2024spectral,atif2024fourier,chen2024physics}. 
Peng et al.\cite{peng2023linear} introduced an attention mechanism within the FNO framework, 
thereby enhancing the prediction accuracy of instantaneous flow field structures. 
Wen et al.\cite{wen2022u} incorporated a U-Net structure into the FNO framework, resulting in 
improved prediction accuracy for the small-scale components of the flow. 
You et al.\cite{you2022learning} introduced implicit recursions into the FNO framework, 
thereby reducing the model's parameter requirements. 
Cao et al.\cite{cao2024spectral} proposed the spectral Fourier neural operator (SFNO), which fine-tunes the 
final linear spectral convolution layer without performing frequency truncation. 
Chen et al.\cite{chen2024physics} proposed a novel physics-enhanced neural operator (PENO) that integrates 
physical knowledge from partial differential equations for accurate predictions, 
while also introducing a self-enhancement mechanism to further reduce cumulative errors 
in long-term simulations. Tran et al.\cite{tran2021factorized} developed a factorized-Fourier neural operator (F-FNO) 
based on the FNO framework, significantly enhancing the prediction accuracy for the 
Navier-Stokes equations.

The FNO methods have been widely applied in turbulence prediction, 
however, their application in chemically reacting turbulent flows still holds significant 
potential. Somdatta et al.\cite{goswami2024learning} applied the deep operator networks (DeepONets) 
method to the prediction of syngas combustion, successfully forecasting the behavior of two-dimensional turbulent jet flames. 
Zhang et al.\cite{zhang2024learning} proposed a multiscale FNO algorithm that learns dynamics of scalar fields across different time scales, successfully predicting the statistics 
of temperature and species in turbulent jet flame fields.
In the prediction of chemically reacting turbulent flows, the impact of swirl motion and strain 
on local scalar structures\cite{hamlington2012intermittency} 
and species mixing, along with the effects of chemical reaction heat release on small-scale flow, 
entropy production statistics, and thermal transport\cite{papapostolou2017enstrophy}
, complicate the direct application of purely data-driven methods.
Instead, it seems to be more effective to establish a mapping from large-scale flow fields 
to subgrid stress (SGS) using neural networks, thereby solving the flow field through 
the LES governing equations\cite{teng2022subgrid}.

Currently, Fourier neural operator-based methods have demonstrated effective applications 
in various turbulent flows, including decaying turbulence\cite{li2023long}
, Rayleigh-Taylor turbulence\cite{luo2024fourier}
, and turbulent channel flow\cite{wang2024prediction}
. This study aims to apply the implicit U-Net enhanced Fourier neural operator to chemically reacting compressible turbulence. 
The rest of this paper is organized as follows:
Section \ref{2} presents the governing equations for chemically reacting compressible turbulence 
in DNS and LES; 
Section \ref{3} introduces the architecture of the Fourier neural operator and the implicit U-Net enhanced Fourier neural operator; 
Section \ref{4} provides the setup of training data and the configuration of IU-FNO; 
Section \ref{5} presents \textit{a posteriori} tests; and finally, 
Section \ref{6} provides a conclusion.

\section{Governing Equations and The Large-Eddy Simulation}\label{2}
This section introduces the governing equations of chemically reacting 
compressible turbulence, followed by LES equations 
and several conventional LES models.

For the chemically reacting compressible turbulence, the three-dimensional 
Navier-Stokes(NS) equations in conservation form are given by\cite{teng2020spectra,teng2022subgrid}:
\begin{equation}
  \frac{\partial\rho}{\partial t}+\frac{\partial\left(\rho u_j\right)}{\partial x_j}=0,
\end{equation}
\begin{equation}
\frac{\partial(\rho u_i)}{\partial t}+\frac{\partial\bigl[\rho u_iu_j+p\delta_{ij}\bigr]}{\partial x_j}=\frac{1}{Re} \frac{\partial\sigma_{ij}}{\partial x_j}+\mathscr{\mathscr{F} }_i,
\end{equation}
\begin{equation}
  \frac{\partial\mathcal{E} }{\partial t}+\frac{\partial\bigl[\bigl(\mathcal{E}+p\bigr)u_j\bigr]}{\partial x_j}=\frac{1}{\alpha}\frac{\partial}{\partial x_j}\left(\kappa\frac{\partial T}{\partial x_j}\right)
  +\frac{1}{Re}\frac{\partial\bigl(\sigma_{ij}u_i\bigr)}{\partial x_j}
  +\Theta-\Lambda+{\mathscr{F} }_ju_j,
\end{equation}
\begin{equation}
  \frac{\partial(\rho Y_s)}{\partial t}+\frac{\partial\big(\rho Y_su_j\big)}{\partial x_j}=\frac{1}{Re} \frac{1}{Sc_s} \frac{\partial}{\partial x_j}\left(\mu\frac{\partial Y_s}{\partial x_j}\right)+\dot{\omega}_s,
  s=1,2,\cdots,n_s-1,
\end{equation}
\begin{equation}
  p=\rho T\big/\big(\gamma M^2\mathcal{M}\big),
\end{equation}
where $\rho$ is the mixture density, $u_i$ is the velocity component, $p$ is the pressure,
 $T$ is the temperature and $\sigma_{ij}$ is the viscous stress, namely,
 \begin{equation}
  \sigma_{ij}=\mu\left(\frac{\partial u_i}{\partial x_j}+\frac{\partial u_j}{\partial x_i}\right)-\frac{2}{3}\mu\theta\delta_{ij},
\end{equation}
where the velocity divergence $\theta=\partial u_k/\partial x_k$ and the 
total energy per unit volume $\mathcal{E}$ is given by 
\begin{equation}
  \mathcal{E}=\frac{p}{\gamma-1}+\frac{1}{2}\rho\big(u_ju_j\big).
\end{equation}
Unless stated otherwise, the summation convention is applied throughout this paper. 
For the chemical reaction component, the source term of the reaction is calculated using 
the Arrhenius-type single-step irreversible reaction equation\cite{teng2020spectra,teng2022subgrid}, namely,
\begin{equation}
  A\rightarrow B,
\end{equation}
where $A$ is the reactant and $B$ is the product. $Y_A$, $Y_B$ and $\omega_A$, $\omega_B$ represent 
the mass fractions and the reaction rates of species $A$ and $B$, respectively. In the whole reaction 
process, we assume that $A$ and $B$ exhibit identical thermodynamic properties, therefore, $\gamma = 1.4$, $C_p$, $C_v$ and the 
mean molecular weight of the mixture $\mathcal{M}$ remains constant throughout the entire reaction process. 
$\mathcal{M}$ is defined as:
\begin{equation}
  \mathcal{M}=1/\sum_{s=1}^{n_s}(Y_s/\mathcal{M}_s),
\end{equation}
where $Y_s$ and $\mathcal{M}_s$ represent mass fraction and dimensionless molecular weight of the sth species, 
respectively. $n_s$ denotes the total number of reactants and products, and in the present reaction, 
$n_s$ equals 2. The reaction rates of reactant ($\dot{\omega}_A$) and product ($\dot{\omega}_B$) are defined as\cite{teng2020spectra,teng2022subgrid},
\begin{equation}
  \dot{\omega}_A=-\dot{\omega}_B=-Da\rho Y_AT^{Ae}\exp\bigl(-Ze/T\bigr).
\end{equation}
The heat source term $\Theta$ is defined as\cite{jaberi1999effects,jaberi2000characteristics}:
\begin{equation}
  \Theta=\frac{\mathrm{Ce}}{(\gamma-1)M^2}\dot{\omega}_B.
\end{equation}
The four reaction-related constants are defined as follows:
\begin{equation}
  Da=2,\quad Ae=1.5,\quad Ze=2,\quad Ce=3.
\end{equation}

The large-scale forcing $\mathscr{F}_i$ is introduced to the solenoidal velocity component by maintaining 
the velocity spectrum fixed\cite{donzis2016statistically} within the two lowest wavenumber bands. The spatially uniform thermal cooling function 
$\Lambda$ is incorporated into the energy conservation equation to partially remove internal energy, thereby balancing 
the energy input from large-scale forcing\cite{wang2010hybrid}.

Large eddy simulation (LES) significantly reduces computational demands compared
 to DNS. The flow field is decomposed into large-scale and small-scale components through
  Favor filtering for compressible turbulence\cite{wang2019cascades,teng2021interscale}, namely,
  \begin{equation}
    \widetilde{g}=\overline{\rho g}/\overline{\rho},
  \end{equation}
  where $\overline{\rho}$ is a filtering operation and defined by\cite{pope2000turbulent}:
  \begin{equation}
    \overline{\rho}(\mathbf{x},t)=\int G(\mathbf{r},\mathbf{x})\rho(\mathbf{x}-\mathbf{r},t)d\mathbf{r}.
  \end{equation}
  The Navier-Stokes equations under the Favor filtering can be reformulated as follows:
  \begin{equation}
    \frac{\partial\overline{\rho}}{\partial t}+\frac{\partial\left(\overline{\rho} \widetilde{u}_j\right)}{\partial x_j}=0,
  \end{equation}
  \begin{equation}
    \frac{\partial(\overline{\rho} \widetilde{u}_i)}{\partial t}+\frac{\partial\left(\overline{\rho} \widetilde{u}_i\widetilde{u}_j+\overline{p}\delta_{ij}\right)}{\partial x_j}-\frac1{Re} \frac{\partial\widetilde{\sigma}_{ij}}{\partial x_j}-\overline{\mathscr{F}_i}= -\frac{\partial\tau_{ij}}{\partial x_j},
  \end{equation}
  \begin{equation}
    \begin{aligned}
      &\frac{\partial\widetilde{\mathcal{E}}}{\partial t}+\frac{\partial\left[\left(\widetilde{\mathcal{E}}+\overline{p}\right)\widetilde{u}_{j}\right]}{\partial x_{j}}-\frac{1}{\alpha}\frac{\partial}{\partial x_{j}}\left(\widetilde{\kappa}\frac{\partial\widetilde{T}}{\partial x_{j}}\right)-\frac{1}{Re}\frac{\partial\left(\widetilde{\sigma}_{ij}\widetilde{u}_{i}\right)}{\partial x_{j}}\\
      &+\overline{\Theta}-\overline{\Lambda}-\overline{\mathscr{F}}_{j}\widetilde{u}_{j}=-\widetilde{u}_{j}\frac{\partial\tau_{ij}}{\partial x_{i}}-\frac{1}{\gamma\left(\gamma-1\right)M^{2}} \frac{\partial Q_{j}}{\partial x_{j}},
    \end{aligned}
  \end{equation}
  \begin{equation}
      \frac{\partial\left(\overline{\rho}\widetilde{Y}_s\right)}{\partial t}+\frac{\partial\left(\overline{\rho}\widetilde{Y}_s\widetilde{u}_j\right)}{\partial x_j}-\frac1{Re}\frac1{Sc_s}\frac\partial{\partial x_j}\left(\widetilde{\mu}\frac{\partial\widetilde{Y}_s}{\partial x_j}\right)-\overline{\dot{\omega}}_s=
      -\frac{\chi_{j,s}}{\partial x_{j}}, s=1,2,\cdots,n_{s}-1.
  \end{equation}
  The effect of small-scale component gives some unclosed subgrid-scale (SGS) terms\cite{mahesh2006large}, including the unclosed SGS stress,
   SGS heat flux, SGS scalar flux and chemical source terms:
  \begin{equation}
    \begin{aligned}
      &\tau_{ij}=\overline{\rho}\left(\widetilde{u_{i}u_{j}}-\widetilde{u_{i}}\widetilde{u_{j}}\right),Q_{j}=\overline{\rho}\left(\widetilde{u_{j}T}-\widetilde{u_{j}}\widetilde{T}\right),
      \chi_{j,s}=\overline{\rho}\left(\widetilde{u_{j}Y_{s}}-\widetilde{u_{j}}\widetilde{Y_{s}}\right),\\&\overline{\dot{\omega}}_{s}=\dot{\omega}_{s}\Big(\overline{\rho},\widetilde{Y}_{s},\widetilde{T}\Big),
      \quad\overline{\Theta}=\Theta\Big(\overline{\rho},\widetilde{Y}_{s},\widetilde{T}\Big).
    \end{aligned}
  \end{equation}
  In the filtered compressible turbulence, the viscous term and the diffusion term are handled in an 
  approximate manner, and the resolved total kinetic energy is defined as,
  \begin{equation}
    \widetilde{\mathcal{E}}=\frac{1}{2}\overline{\rho}(\widetilde{u}_{j}\widetilde{u}_{j}) + \overline{p}/(\gamma-1).
  \end{equation}

Traditional SGS models include the velocity gradient model, the Smagorinsky model (SM), 
dynamic Smagorinsky model (DSM), and dynamic mixed model (DMM). In this paper, we select the DSM 
approach as the subgrid scale modeling method for large eddy simulation (LES). This model 
assumes that the subfilter scale stress follows a similar constitutive equation to that of molecular 
viscous stress. The DSM can be expressed as\cite{smagorinsky1963general}:
\begin{equation}\begin{aligned}
  &\tau_{ij}^{DSM}-\frac{\delta_{ij}}{3}\tau_{kk}^{\mathrm{DSM}}=-2C_{s}^{2}\Delta^{2}|\widetilde{S}|\left(\widetilde{S}_{ij}-\frac{\delta_{ij}}{3}\widetilde{S}_{kk}\right), \\
  &\tau_{kk}^{DSM}=2C_{I}\Delta^{2}|\widetilde{S}|^{2}, \\
  &Q_{j}^{DSM}=-C_{T}\Delta^{2}|\widetilde{S}|\frac{\partial\widetilde{T}}{\partial x_{j}}, \\
  &\chi_{j,s}^{DSM}=-C_{T}\Delta^{2}|\widetilde{S}|\frac{\partial\widetilde{Y}_{s}}{\partial x_{j}} .
  \end{aligned}\end{equation}

Here, the characteristics stain rate $|\widetilde{S}| = \sqrt{2\widetilde{S}_{ij}\widetilde{S}_{ij}}.$ 
Based on the Germano identity\cite{xie2020approximate,xie2020spatial,xie2020spatially}, 
the model coefficients $C_s^2$, $C_I$ and $C_T$ can be dynamically determined by the least-squares method. 

\section{The Fourier neural operator and the implicit U-net enhanced 
fourier neural operator}\label{3}

The Fourier neural operator (FNO) offers several advantages in learning turbulent flow fields. 
Compared to traditional neural networks, FNO leverages the Fourier space to learn the 
target set, thereby achieving a better prediction for multi-scale flow field characteristics. 
Additionally, during the learning process, FNO applies truncation operations to 
filter out small-scale noise in the flow field for enhancing computational efficiency\cite{li2022fourier}. In this section, the network structures of FNO and IU-FNO 
are presented.
\begin{figure}[htbp]
  \centering
  \includegraphics[width=\textwidth]{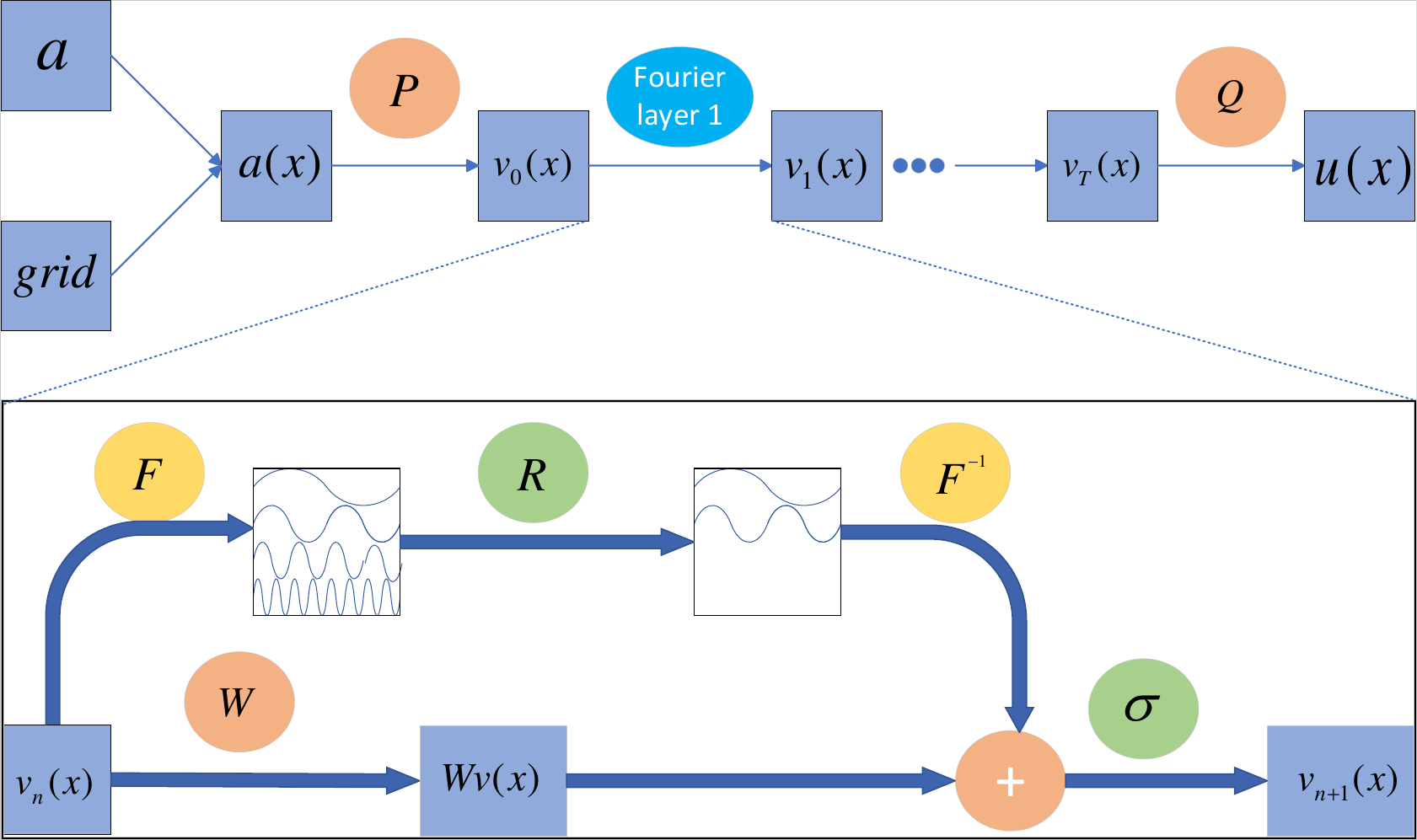} 
  \caption{The configuration of Fourier neural operator} 
  \label{fno} 
\end{figure}

\subsection{\label{sec:level2}The Fourier neural operator}
FNO is designed to map between two infinite-dimensional space by continuously 
learning and optimizing the operator $G$, namely,
\begin{equation}
  G:\mathcal{A}\times\Theta\to\mathcal{U}\quad\text{ or }\quad G_\theta:\mathcal{A}\to\mathcal{U},\quad\theta\in\Theta, 
\end{equation}
where $\mathcal{A}=\mathcal{A}(D;\mathbb{R}^{d_{a}})$ and 
$\mathcal{U}=\mathcal{U}(D;\mathbb{R}^{d_{u}})$ take values $\mathbb{R}^{d_{a}}$ and
$\mathbb{R}^{d_{u}}$ in separable Banach spaces, respectively\cite{li2020fourier}. The parameter set 
$\theta$ is iteratively optimized through the FNO network. The architecture of FNO is displayed
in Fig.~\ref{fno} and detailed as follows: 

(1) The input dataset $a$ is concatenated with the grid dataset 
$g(x)$ to form the input $a(x)$, where $g(x)$ is generated from a sequence of spatial coordinates.

(2) Input $a(x) \in \mathcal{A} $ is mapped from a low-dimensional space to a high-dimensional 
space through the local transformation $P$, namely $v_{0}(x)=P(a(x))$, thereby 
increasing the number of parameters for operator learning. 
Specifically, $P$ is a single-layer fully connected neural network.

(3) The high-dimensional representation $v_{0}(x)$ can undergo multiple iterations 
by the Fourier layers, which can be written as:
\begin{equation}
  v_{t+1}(x):=\sigma\Big(Wv_t(x)+\Big(\mathcal{K}(a;\phi)v_t\Big)(x)\Big),\quad\forall x\in D.
\end{equation}
The iterative process primarily consists of three components, namely 
the linear transformation component $Wv_{t}(x)$,  the non-local integral operator component 
$\big(\mathcal{K}(a;\phi)v_{t}\big)(x)$ and non-linear activation function $\sigma:\mathbb{R}\to\mathbb{R}$, 
where the kernel 
$\mathcal{K}:\mathcal{A}\times\Theta_{\mathcal{K}}\to\mathcal{L}(\mathcal{U}(D;\mathbb{R}^{d_{v}}), \mathcal{U}(D;\mathbb{R}^{d_{v}}))$
maps to bounded linear operators on $\mathcal{U}(D;\mathbb{R}^{d_{v}})$ and is parameterized
by $\phi\in\Theta_{\mathcal K}$.

To define the Fourier integral operator $\mathcal{K}$, firstly we denote the Fourier 
transform $\mathcal{F}$ and its inverse $\mathcal{F}^{-1}$ as,  
\begin{equation}
  \begin{aligned}
  (\mathcal{F}f)_j(k)=\int_Df_j(x)e^{-2i\pi\langle x,k\rangle}\mathrm{d}x,
  \quad(\mathcal{F}^{-1}f)_j(x)=\int_Df_j(k)e^{2i\pi\langle x,k\rangle}\mathrm{d}k,
  \end{aligned}
\end{equation}
and $\mathcal{K}$ in Fourier space can be written as: 
\begin{equation}
  \big(\mathcal{K}(\phi)v_t\big)(x)=\mathcal{F}^{-1}\Big(R_\phi\cdot(\mathcal{F}v_t)\Big)(x)\quad\forall x\in D,
\end{equation}
where $R_{\phi}$ is the Fourier transform of a periodic function $\kappa : \bar{D} \to \mathbb{R}^{d_{v}\times d_{v}}$
parameterized by $\phi\in\Theta_{\mathcal{K}}$. For the Fourier integral operator $\mathcal{K}$, 
we can choose the appropriate $k_{\mathrm{max}}$ as the maximal number of mode to 
truncate the Fourier series. Here $k_{\max}=|Z_{k_{\max}}| = |\{k \in \mathbb{Z}^{d} : |k_{j}| \leq k_{\max,j}, \mathrm{for} j = 1,\ldots,d\}|$,
and thus $R_{\phi}$ is parameterized as complex-valued-tensor $(k_{\max}\times d_{v}\times d_{v})$
and comprising a collection of truncated Fourier modes\cite{li2020fourier}. In practical computations, we apply 
fast Fourier transform (FFT). Therefore, the Fourier truncation operation can be written as:
\begin{equation}
  \begin{aligned}
  \left(R\cdot(\mathcal{F}v_t)\right)_{k,l}=\sum_{j=1}^{d_v}R_{k,l,j}(\mathcal{F}v_t)_{k,j},&\quad k=1,\ldots,k_{\max},
  \quad j=1,\ldots,d_v.
  \end{aligned}
\end{equation}
where the FFT $\hat{\mathcal{F}}$ and its inverse $\hat{\mathcal{F}}^{-1}$ are defined as:
\begin{equation}
  \begin{aligned}
    &(\hat{\mathcal{F}}f)_l(k)=\sum_{x_1=0}^{s_1-1}\cdots\sum_{x_d=0}^{s_d-1}f_l(x_1,\ldots,x_d)e^{-2i\pi\sum_{j=1}^d\frac{x_jk_j}{s_j}},\\
    &(\hat{\mathcal{F}}^{-1}f)_l(x)=\sum_{k_1=0}^{s_1-1}\cdots\sum_{k_d=0}^{s_d-1}f_l(k_1,\ldots,k_d)e^{2i\pi\sum_{j=1}^d\frac{x_jk_j}{s_j}},\\
    &l = 1,...,d_v.
  \end{aligned}
\end{equation}

(4) After passing through the final Fourier layer, the resulting high-dimensional 
representation is processed via the local transformation $Q$, yielding the final 
target output $u(x)$, where $u(x)=Q(v_{T}(x))$. Here, $Q$ is a 
single-layer fully connected neural network\cite{li2020fourier}.

\begin{figure}[htbp]
  \centering
  \includegraphics[width=\textwidth]{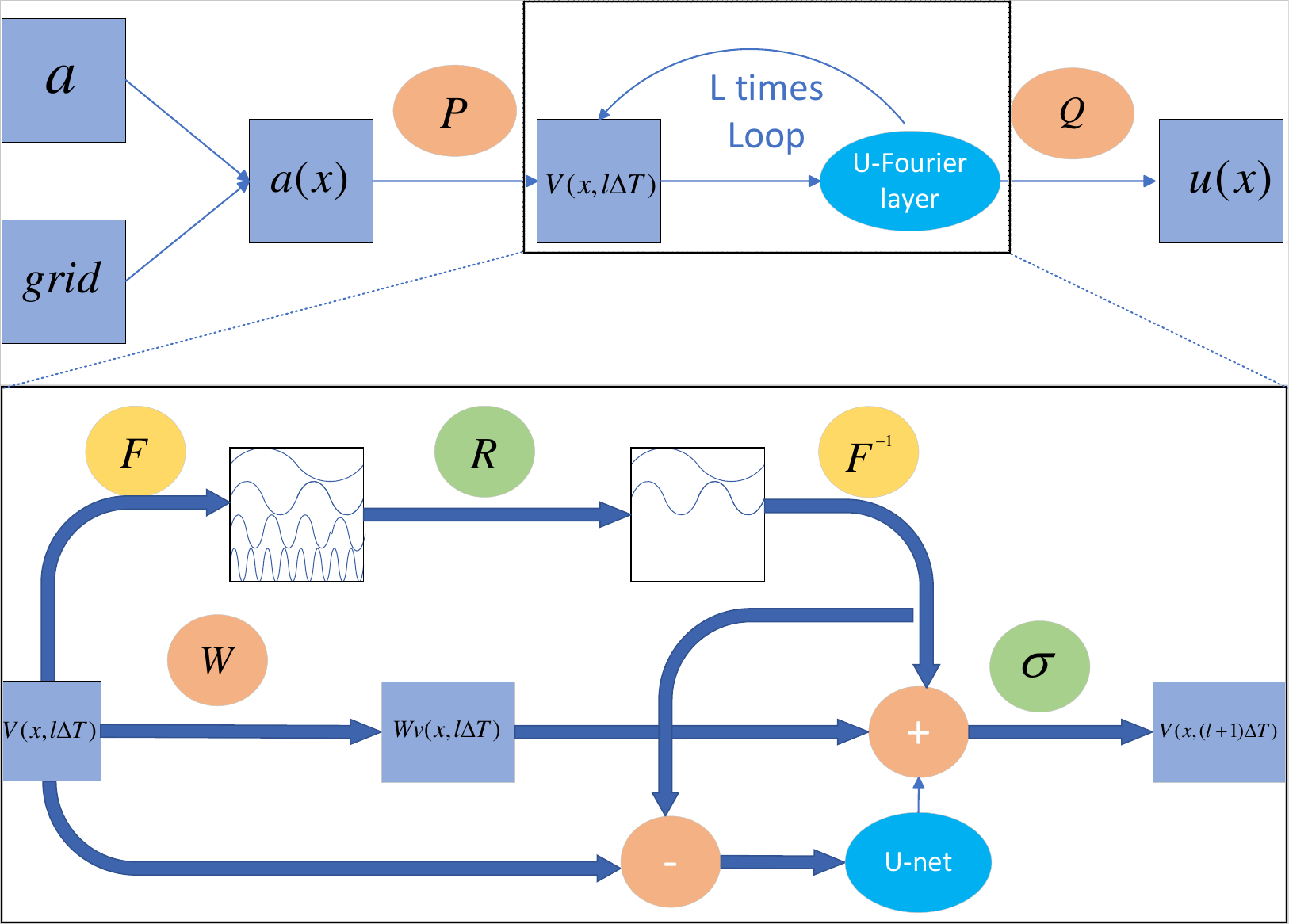} 
  \caption{The configuration of implicit U-Net enhanced Fourier neural operator} 
  \label{iu} 
\end{figure}

\subsection{\label{sec:level2}The implicit U-Net enhanced Fourier neural operator}
The implicit U-Net enhanced Fourier neural operator (IU-FNO) combines the features of both IFNO and U-FNO, and its configuration is 
presented in Fig.~\ref{iu}. The primary difference between IU-FNO and FNO lies in the Fourier 
layer. While FNO employs multiple Fourier layers in sequence, IU-FNO utilizes 
a shared Fourier layer, where a single Fourier layer parameters is reused\cite{li2023long}. 
Additionally, IU-FNO incorporates the U-net structure within the Fourier layer, 
enabling the model to capture and learn small-scale information more effectively\cite{ronneberger2015u,wen2022u}.

The iterative process of the IU-FNO network can be expressed by the following equation:
\begin{equation}
  v(x,(l+1)\Delta t)=\mathcal{L}^\text{IUFNO}[v(x,l\Delta t)]:=v(x,l\Delta t)+\Delta t\sigma\left(c(x,l\Delta t)\right),\quad\forall x\in D,
\end{equation}
where, $\Delta t$ represents the implicit iteration step size within the Fourier layer, 
$l$ denotes the number of iterations, $\sigma$ is the activation function, 
and $c$ is defined by:
\begin{equation}
  c(x,l\Delta t):=Wv(x,l\Delta t)+\mathcal{F}^{-1}\left(R_\phi\cdot(\mathcal{F}v(x,l\Delta t))\right)(x)+\mathcal{U}^*s(x,l\Delta t),\quad\forall x\in D,
\end{equation}
where the internal parameter operations of the Fourier layer are specifically illustrated.
Similar to FNO, $Wv(x,l\Delta t)$ and $\mathcal{F}^{-1}(R_\phi\cdot(\mathcal{F}v(x,l\Delta t)))(x)$
represent the linear transformation and Fourier truncation, respectively, while $\mathcal{U}^*s(x,l\Delta t)$ represents 
the U-Net which is denoted by $\mathcal{U}^*$ and capture of the residual field. The residual field can be defined as\cite{li2023long}:
\begin{equation}
  s(x,l\Delta t):=v(x,l\Delta t)-\mathcal{F}^{-1}\left(R_\phi\cdot\left(\mathcal{F}v(x,l\Delta t)\right)\right)(x),\quad\forall x\in D.
\end{equation}

By introducing implicit iteration and the U-net architecture, IU-FNO is able to reduce 
the number of parameters while simultaneously improving the overall accuracy of the model.\cite{li2022fourier}

\section{the dns dataset description and IU-FNO model settings}\label{4}
In previous section, the network structures of FNO and IU-FNO are introduced. 
In this section, we describe the generation of filtered DNS data for three-dimensional 
chemically reacting compressible turbulence, which serve as the input-output pairs 
for IU-FNO. We also introduce the configuration of the IU-FNO network parameters.

\begin{figure}[h]
  \centering
  \begin{subfigure}[b]{0.49\textwidth}
      \centering
      \includegraphics[width=\textwidth]{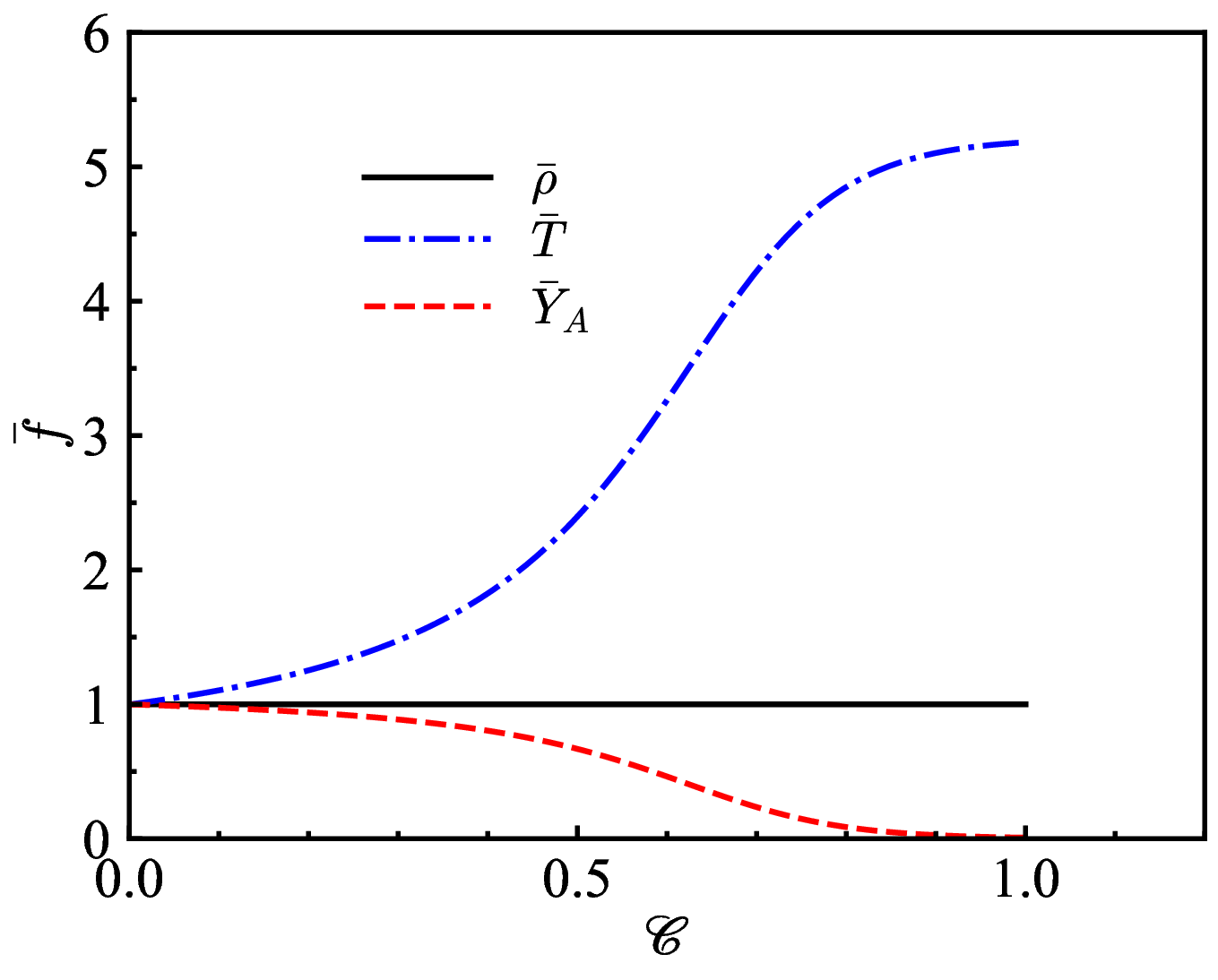}  
      \caption{}  
      \label{mean}
  \end{subfigure}
  \hfill
  \begin{subfigure}[b]{0.49\textwidth}
      \centering
      \includegraphics[width=\textwidth]{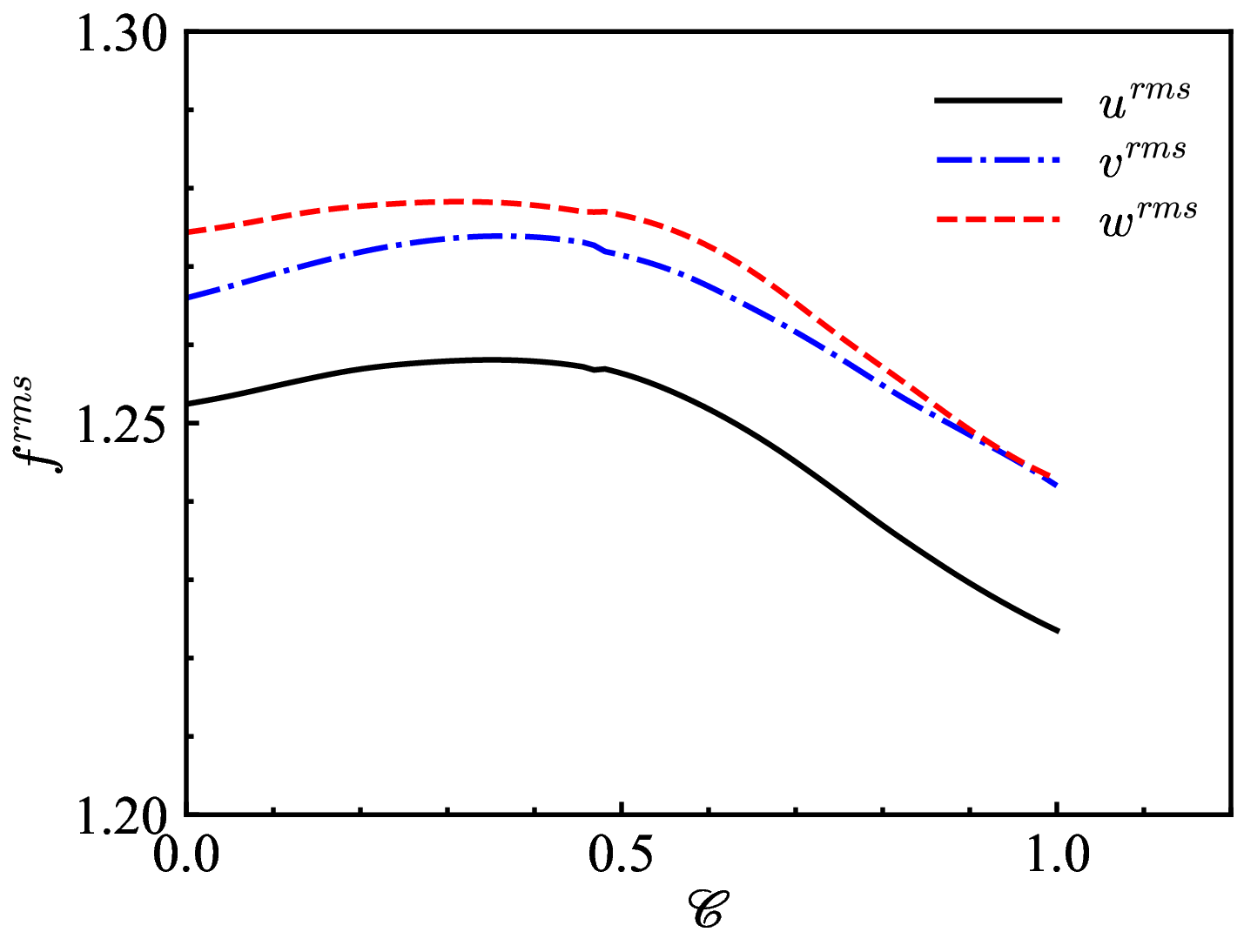}  
      \caption{}  
      \label{rms1}
  \end{subfigure}
  \begin{subfigure}[b]{0.49\textwidth}
    \centering
    \includegraphics[width=\textwidth]{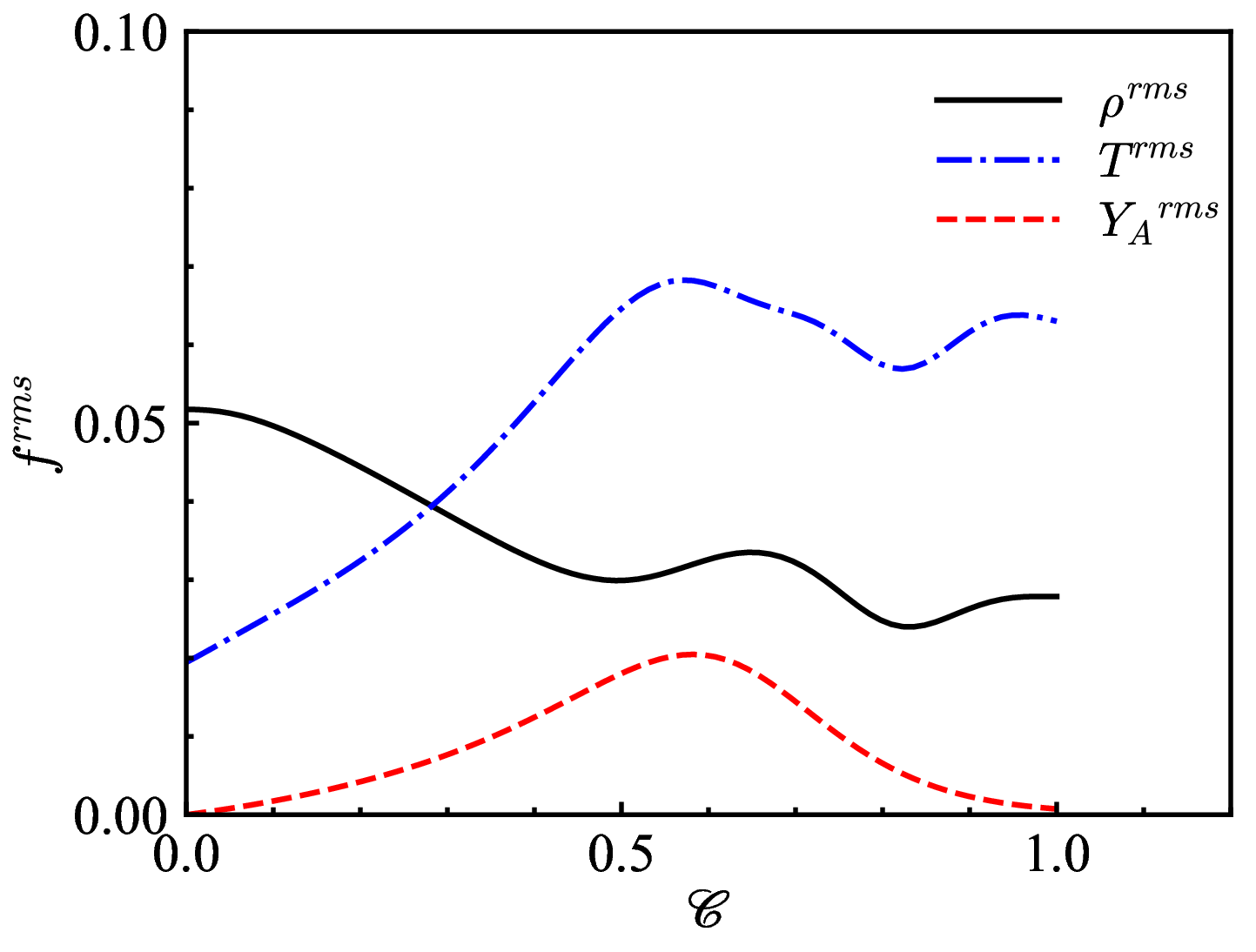}  
    \caption{}  
    \label{rms2}
\end{subfigure}
  \caption{Time evolution of the spatial average values $\bar{f}$ and the rms values $f^{rms}$. 
  (a) The spatial average values of $\rho$, $T$ and $Y_A$, (b) the rms values of $u_i$ and (c) the rms values of $\rho$, $T$ and $Y_A$}
  \label{mean and rms}
\end{figure}
\subsection{\label{sec:level2}The DNS dataset description}
To generate the three-dimensional chemically reacting compressible turbulence 
dataset for IU-FNO training and testing, we set the grid resolution to $256^{3}$ 
and apply periodic boundary conditions in all three spatial directions. 
Numerical simulations are conducted within a cubic domain of $(2\pi)^3$.\cite{teng2020spectra}

The reaction progress variable $\mathscr{C}$ defined by $\mathscr{C}=(T-T_{int})/(T_{ad}-T_{int})$
is used to evaluate the chemically reacting process\cite{nikolaou2019progress}. 
Here, $T$ represents the dimensionless temperature of the mixture, $T_{int}$ is the initial 
dimensionless temperature of the mixture before the start of chemical reaction and $T_{ad}$ 
is the dimensionless mixture temperature at the end of chemical reaction ideally for $Y_A < 0.01$ undergoing 
an adiabatic process. In the reaction used in our study, $T_{int}=1.0\mathrm{~and~}T_{ad}=5.18$. 
Throughout the reaction process, $\mathscr{C}$ progresses from 0 to 1.


\begin{table}
  \caption{Statistics of DNS database at the different $\mathscr{C}$ (N = $256^3$).\label{statistics}}
  \begin{ruledtabular}
  \begin{tabular}{cccccccccc}
  $\mathscr{C}$ & $M_t$ & $Re_\lambda$ & $\eta/\Delta_{DNS}$ & $L_{I/\eta}$ & $\lambda/\eta $ & $Y_A$ & $\langle T\rangle $ & $E_K$& $\epsilon$\\
  \hline
  0 & 0.4 & 100 & 1 & 63 & 20 & 1 & 1 & 2.60 & 0.72\\
  1 & 0.173 & 45 & 2.3 & 29 & 15 & 0.001 & 5.18 & 2.51 & 0.65\\
  \end{tabular}
  \end{ruledtabular}
  \end{table}

Table~\ref{statistics} presents various statistical quantities at the initial and final 
stages of the reaction.The Taylor microscale Reynolds number $Re_\lambda $ and 
the turbulent Mach number $M_t$ are defined, respectively by\cite{wang2012effect,wang2010hybrid}:
\begin{equation}
  Re_\lambda=Re\frac{\langle\rho\rangle u^{'}\lambda}{\sqrt{3}\langle\mu\rangle},\quad M_t=M\frac{u^{'}}{\langle\sqrt{T}\rangle},
\end{equation}
where, $\left\langle\cdot\right\rangle $ stands for spatial average. The Taylor microscale is: 
\begin{equation}
  \lambda=\sqrt{\frac{\langle u_1^2+u_2^2+u_3^2\rangle}{\left\langle\left(\partial u_1/\partial x_1\right)^2+\left(\partial u_2/\partial x_2\right)^2+\left(\partial u_3/\partial x_3\right)^2\right\rangle}}.
\end{equation}
$L_{I}$ and $\eta$ represent the integral length scale and the Kolmogorov ength scale respectively\cite{wang2012effect,wang2010hybrid} 
and are defined by:
\begin{equation}
  \begin{aligned}
  &L_I=\frac{3\pi}4\frac{\int_0^\infty[E(k)/k]dk}{\int_0^\infty E(k)dk},\\
  &\eta=\left[\left\langle\mu/\rho\right\rangle^3/\epsilon\right]^{1/4}.
  \end{aligned}
\end{equation}
$E_{K}$ is the energy spectrum and defined by\cite{fan2023eddy,pope2000turbulent,wang2012effect}: 
\begin{equation}
    \int_0^\infty E(k) \mathrm{d}k=(u^{\prime})^{2}/2, u^{\prime}\equiv\sqrt{\langle u_1^2+u_2^2+u_3^2\rangle},
\end{equation}
$\epsilon=\frac{1}{Re}\left\langle\frac{\sigma_{ij}}{\rho}\frac{\partial u_i}{\partial x_j}\right\rangle $ 
stands for kinetic energy dissipation. 

As observed from the table~\ref{statistics}, during the process of the reaction, the turbulent 
Mach number decreases from 0.4 to 0.17. Simultaneously, the Taylor Reynolds number 
decreases due to the increase in viscosity caused by the exothermic chemical reaction. 
Throughout the reaction, both turbulent kinetic energy and kinetic energy dissipation 
experience a decline.

To train the IU-FNO model, we prepare 400 sets of chemically reacting turbulence 
data with different initial conditions. Each dataset comprises 7800 DNS time steps, 
spanning from the beginning of the reaction $(\mathscr{C}=0)$ to the end of the reaction
$(\mathscr{C}=1)$\cite{teng2022subgrid}. The DNS time step is defined as 1/10000 of the large eddy 
turnover time $\tau$, where $\tau$=1.2 in our study. To reduce the computational cost 
of training the neural network, the chemically reacting turbulence data, originally at 
$256^3$ grid resolution, are downsampled to $32^3$ resolution by skipping intervals, 
resulting in filtered DNS (fDNS) data. This downsampled data serves as the training data  
for our study. 

During chemically reacting process, the spatial average values of the velocity components in 
three directions remain nearly constant and are very close to zero. The spatial average 
values of other important physical quantities
are shown in Fig.~\ref{mean}. 
We can see that the spatial averages
of the dimensionless temperature and mass fraction of $Y_A$ change over time as 
the reaction progresses, while the spatial average density remains nearly constant.
During the reaction, the average temperature $T$ increases from 1 to 5.18, while the average reactant 
mass fraction $Y_A$ decreases from 1 to 0.01. This variation continuously impacts the 
backpropagation process of the IU-FNO model. At the beginning of the reaction, the 
weight of temperature in the loss function is relatively low, whereas by the end of 
the reaction, the weight of temperature in the loss error becomes higher. In contrast, 
the weight of the mass fractions of reactant follows the opposite trend. Additionally, 
throughout the reaction, the root mean square (rms) value of the velocity is much larger 
than those of the density, temperature, and component fields, shown in Figs.~\ref{rms1} and \ref{rms2}, 
indicating that the velocity 
field exhibits larger spatial fluctuations. To ensure that the 
influence of each physical quantity on the weights of model is balanced, it is common 
to adjust the loss function or apply data preprocessing methods. In this case, we 
use data preprocessing. Using the function defined as, 

\begin{equation}
  f^*=\frac{f-\bar{f}}{f^{rms}}
\end{equation}
each physical quantity $f$ is normalized to improve the model's accuracy\cite{luo2024fourier}.

\subsection{\label{sec:level2}IU-FNO model settings}
\begin{figure}[h]
  \centering
  \begin{subfigure}[b]{0.49\textwidth}
      \centering
      \includegraphics[width=\textwidth]{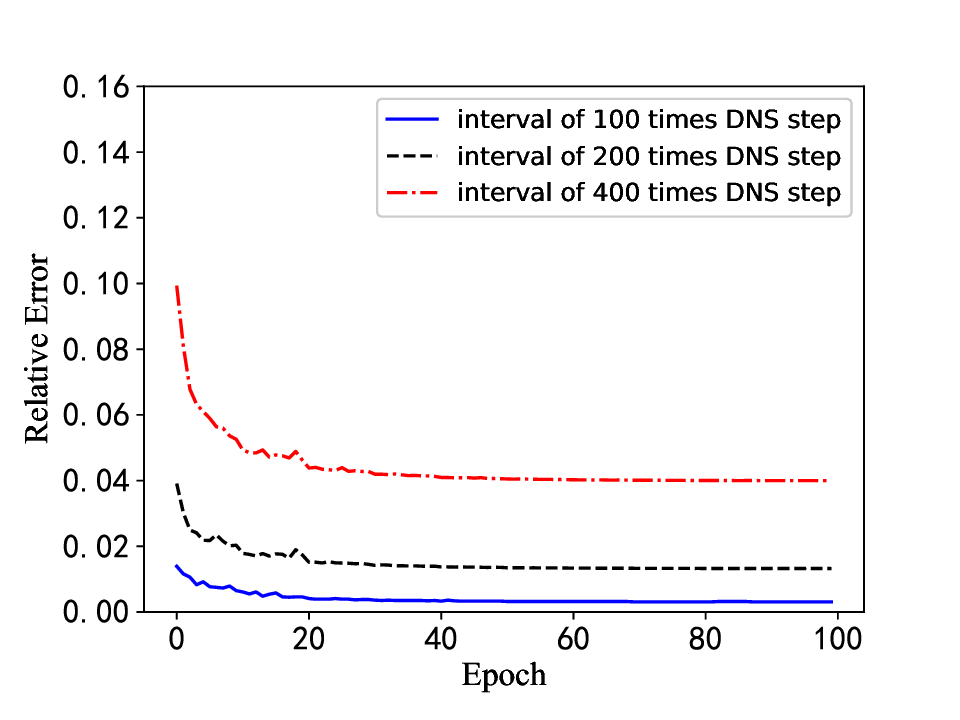}  
      \caption{}  
      \label{loss1}
  \end{subfigure}
  \hfill
  \begin{subfigure}[b]{0.49\textwidth}
      \centering
      \includegraphics[width=\textwidth]{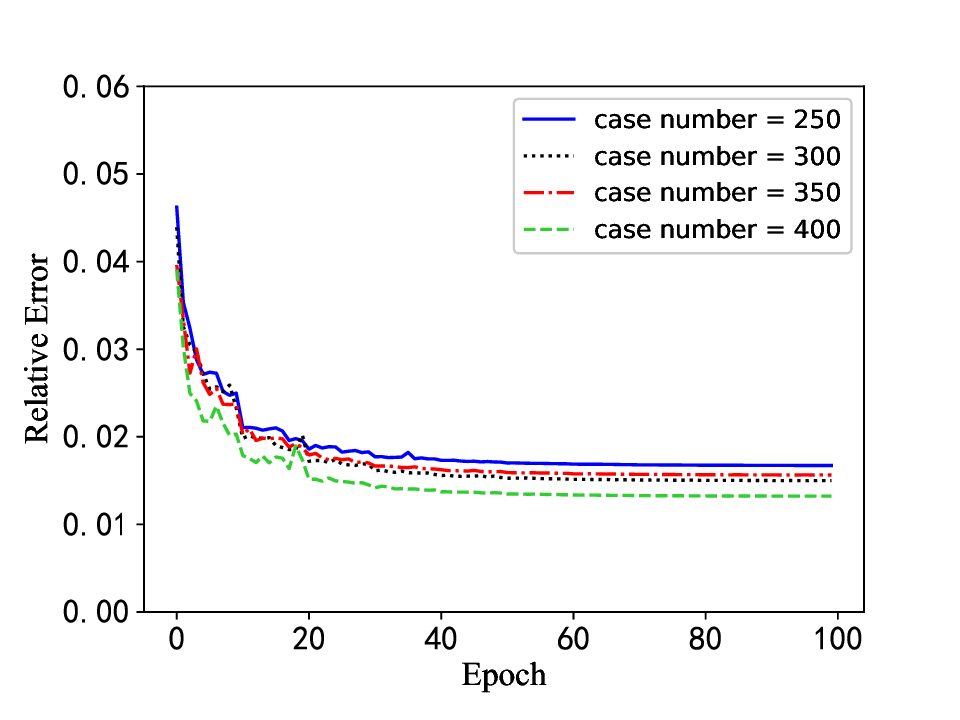}  
      \caption{}  
      \label{loss2}
  \end{subfigure}
  \begin{subfigure}[b]{0.49\textwidth}
    \centering
    \includegraphics[width=\textwidth]{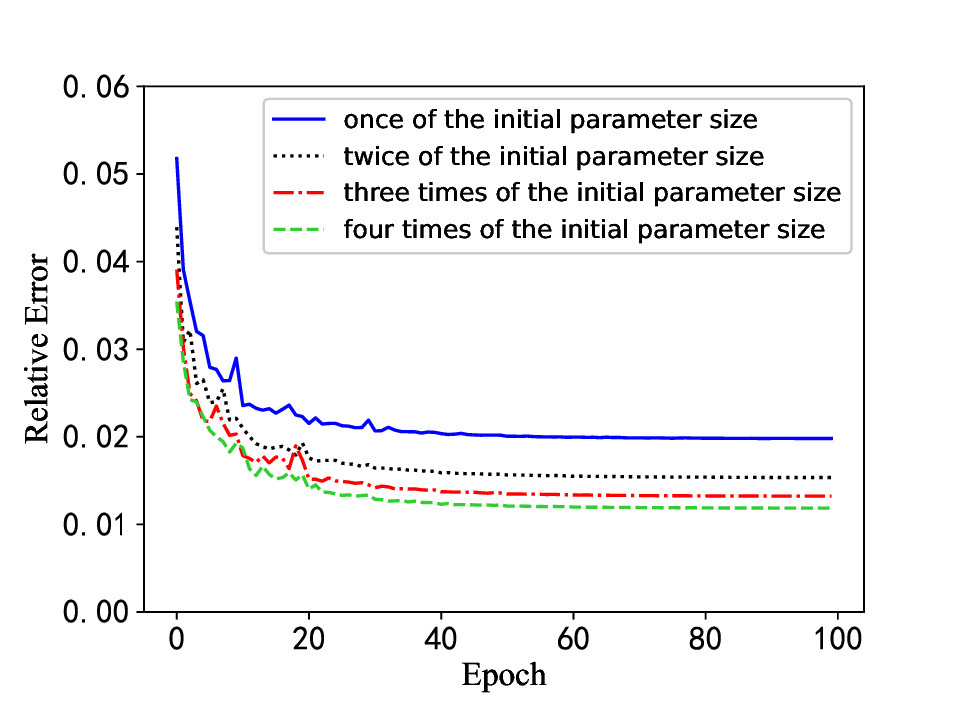}  
    \caption{}  
    \label{loss3}
  \end{subfigure}
  \begin{subfigure}[b]{0.49\textwidth}
    \centering
    \includegraphics[width=\textwidth]{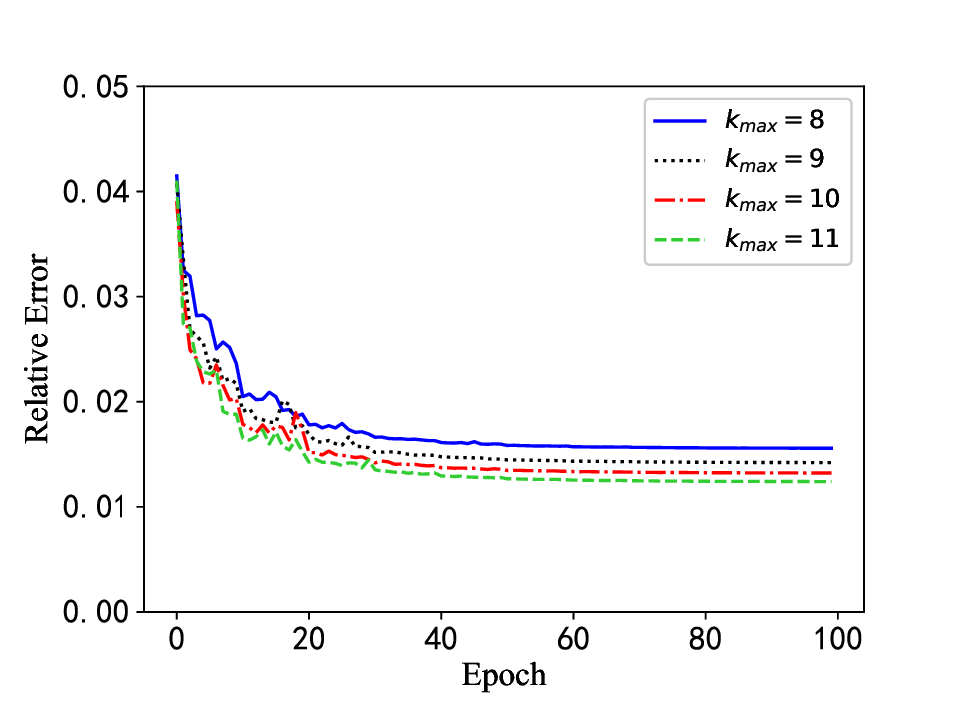}  
    \caption{}  
    \label{loss4}
  \end{subfigure}
 \caption{The evolutions of the testing loss with different parameter. (a) interval of DNS data for training (b) amounts of DNS data for training (c) channel width (d) truncation wavenumber.}
  \label{loss}
\end{figure}
We use $[U_{m-4},U_{m-3},U_{m-2},U_{m-1},U_m]$
as the input to the network of IU-FNO and $U_{m+1}$ is output, where $U_{n}$ represents the 
physical field information at the n-th time step, including the density $\rho$, 
the three velocity components $(u, v, w)$, temperature $T$, and the mass fraction $Y_A$.
The data shape of $U$ is $[N_x,N_y,N_z,n]$, where $N_x=32$, $N_y=32$ and $N_z=32$ 
represent the numbers of grid points in three coordinate directions, and 
$n=6$ corresponds to the number of physical quantities. The training and testing 
losses are defined as\cite{peng2022attention,li2022fourier}: 
\begin{equation}
  \mathrm{Loss}=\frac{\| \mathbf{f}^*- \mathbf{f}\|_2}{\| \mathbf{f}\|_2},\quad\mathrm{where}\quad\|\mathbf{x}\|_2=\frac1N\sqrt{\sum_{k=1}^N|x_k|^2}.
\end{equation}
Here, $ \mathbf{f}^{*}$ denotes the prediction of density, velocity, temperature and mass fraction of $Y_A$ increments, and $ \mathbf{f}$ is the 
ground truth. For the optimizer and activation function, we select the Adam optimizer\cite{kingma2014adam} 
and the GELU activation function\cite{hendrycks2016gaussian} respectively, with the initial learning rate setting to $10^{-3}$.

Due to the small DNS time step, 
the differences between consecutive time steps are minimal. If DNS time steps are 
directly used as input, IU-FNO may fail to capture meaningful physical information. 
Therefore, we generally sample data at intervals spanning several DNS time steps to train 
the IU-FNO. We test training data with 
intervals of 100, 200, and 400 DNS steps, as shown in Fig.~\ref{loss1}. It can be observed 
that when the interval is 400 times DNS steps, the corresponding test error is significantly 
larger 
compared to the interval of 200 times DNS steps, while the test errors for the 
intervals of 100 and 200 times DNS steps are both small. Due to the error accumulation effect in time 
iteration 
when using IU-FNO, we aim to minimize the number of time steps required, within 
the allowable error range. Therefore, in our study, we selected an interval of 
200 DNS time steps to obtain the training data. 

Regarding the amount of training 
data, we test the impact of different training set sizes on the test error. Specifically, 
we test the case 
number of 250, 300, 350, 400 and record the relative error across epochs in Fig.~\ref{loss2}. 
Here, case number refers to the number of initial conditions of chemically reacting 
turbulence fields used for training the model, and each of these initial conditions 
is generated randomly. As the size of the training 
set increases, the training error consistently decreases. When the number of cases 
exceeds 300, the test error no longer significantly decreases with the addition of 
more training data, indicating that with 400 cases, the dataset is sufficient for IU-FNO 
to capture dynamics of chemically reacting turbulence 
corresponding to different initial conditions. Therefore, we 
select 400 randomly initialized chemically reacting processes for the training set, with 
each case containing 7800 DNS steps. With a 200-step interval, each case yields 39 data 
points. Since IU-FNO uses the first five steps to predict the sixth, the actual number of 
input-output pairs is $34\times400=13,600$. Moreover, additional five independently generated cases are used for 
\textit{a posteriori} tests.

When selecting the parameters for the IU-FNO network, we test the different channel widths 
in the fully connected operation $P$, 
as well as the Fourier truncation wavenumber $k_{max}$ in the operation $R$.
For the channel width in the fully connected operation $P$, 
we selected once, twice, three times, and four times of the initial parameter size where the initial parameter size equals to 33, 
while for the truncation wavenumber, we tested $k_{max}=$8, 9, 10 and 11. Figs.~\ref{loss3} and~\ref{loss4} shows the impact of different channel widths 
and truncation 
wavenumbers on the IU-FNO test error. After considering both training accuracy and 
computational cost, we selected a channel width three times of the initial parameter size and a truncation wavenumber
of $k_{max}=10$ as the parameters for the Fourier layer. 

\section{\textit{a posteriori} test}\label{5}
In the \textit{a posteriori} tests, we utilize five sets of three-dimensional fDNS initial fields that are independent from the 
training set to evaluate the predictive performance of IU-FNO. We also apply traditional LES 
with DSM model for comparative analysis, with a grid resolution of $32^3$. 
The approach involves 
inputting the initial flow field into the trained model for prediction and iteratively advancing the prediction in time until the 
reaction is almost complete, specifically at $\mathscr{C}=1$. All statistical results are ensemble averaged and compared with 
those from fDNS and DSM.
\begin{figure}[htbp]
  \centering
  \begin{subfigure}[b]{0.4\textwidth}
      \centering
      \includegraphics[width=\textwidth]{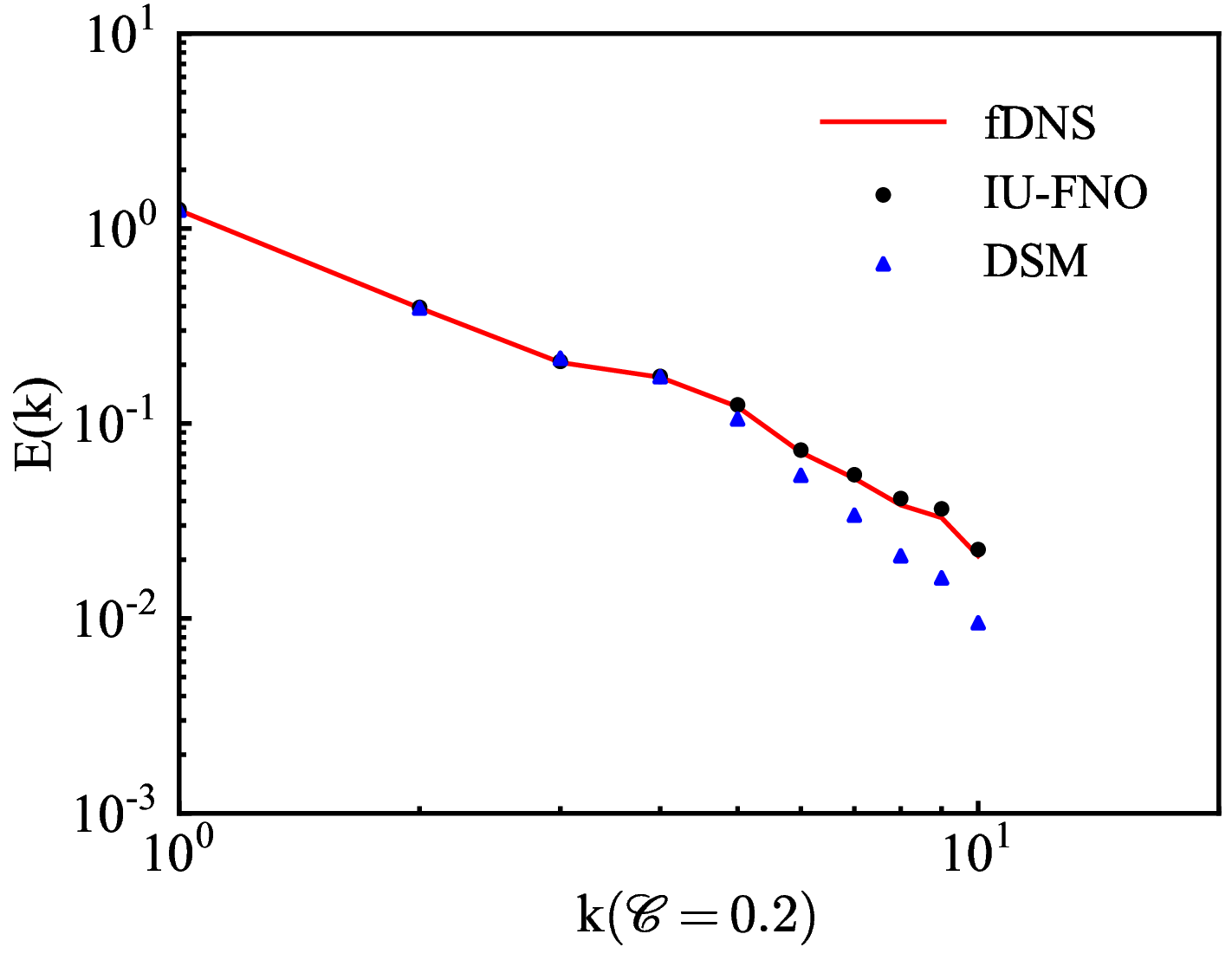}  
      \caption{}  
      \label{spectral1}
  \end{subfigure}
  \begin{subfigure}[b]{0.4\textwidth}
      \centering
      \includegraphics[width=\textwidth]{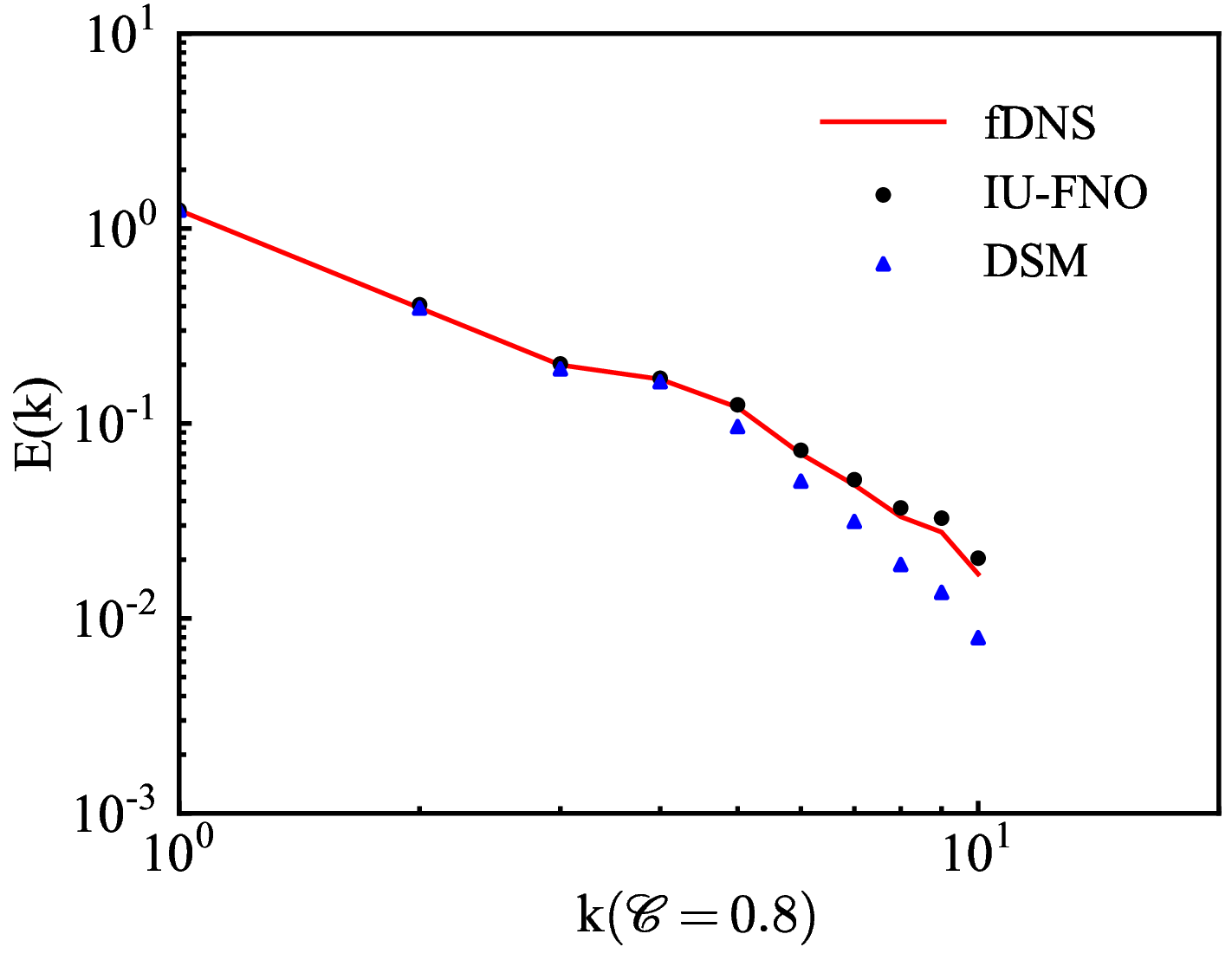}  
      \caption{}  
      \label{spectral2}
  \end{subfigure}
  \begin{subfigure}[b]{0.4\textwidth}
    \centering
    \includegraphics[width=\textwidth]{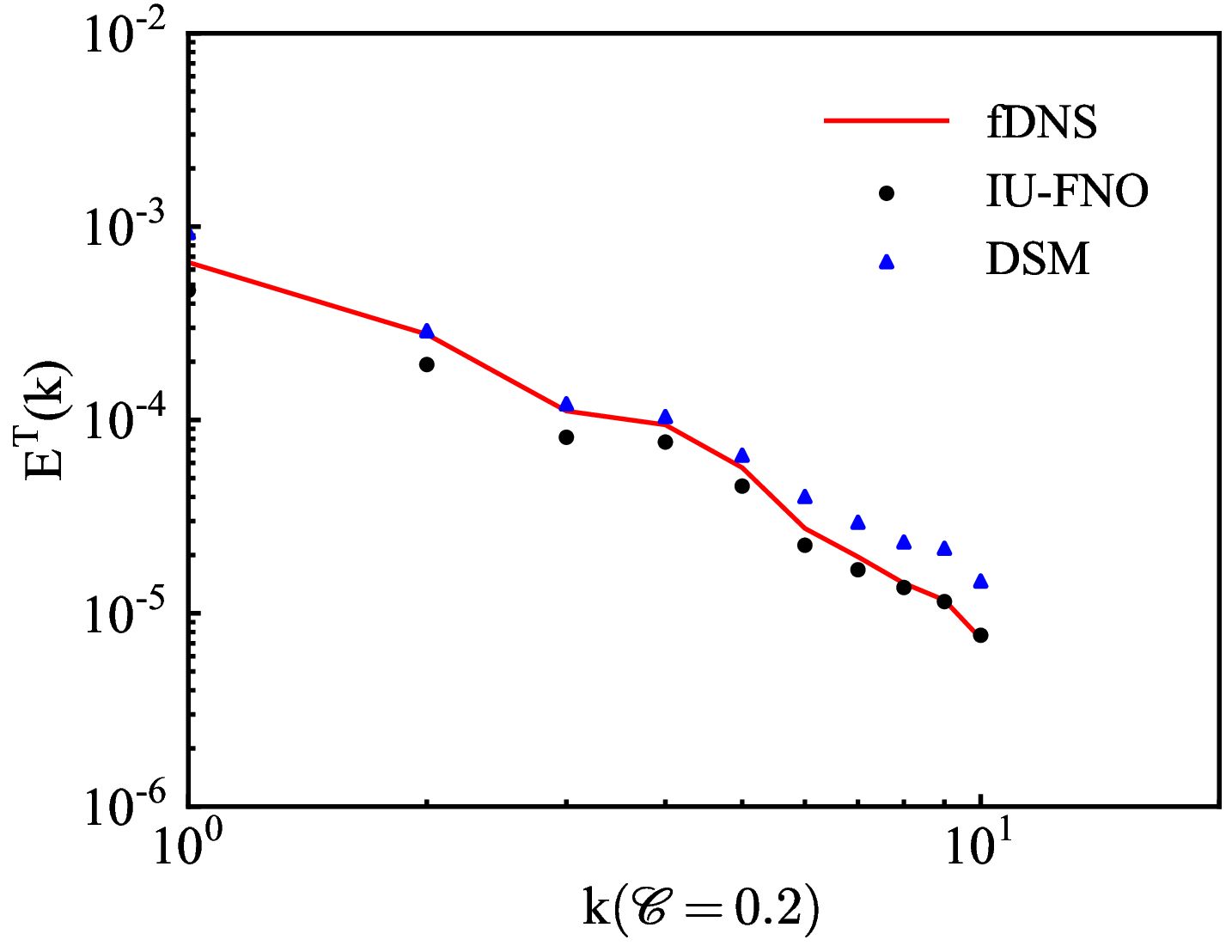}  
    \caption{}  
    \label{spectral3}
  \end{subfigure}
  \begin{subfigure}[b]{0.4\textwidth}
    \centering
    \includegraphics[width=\textwidth]{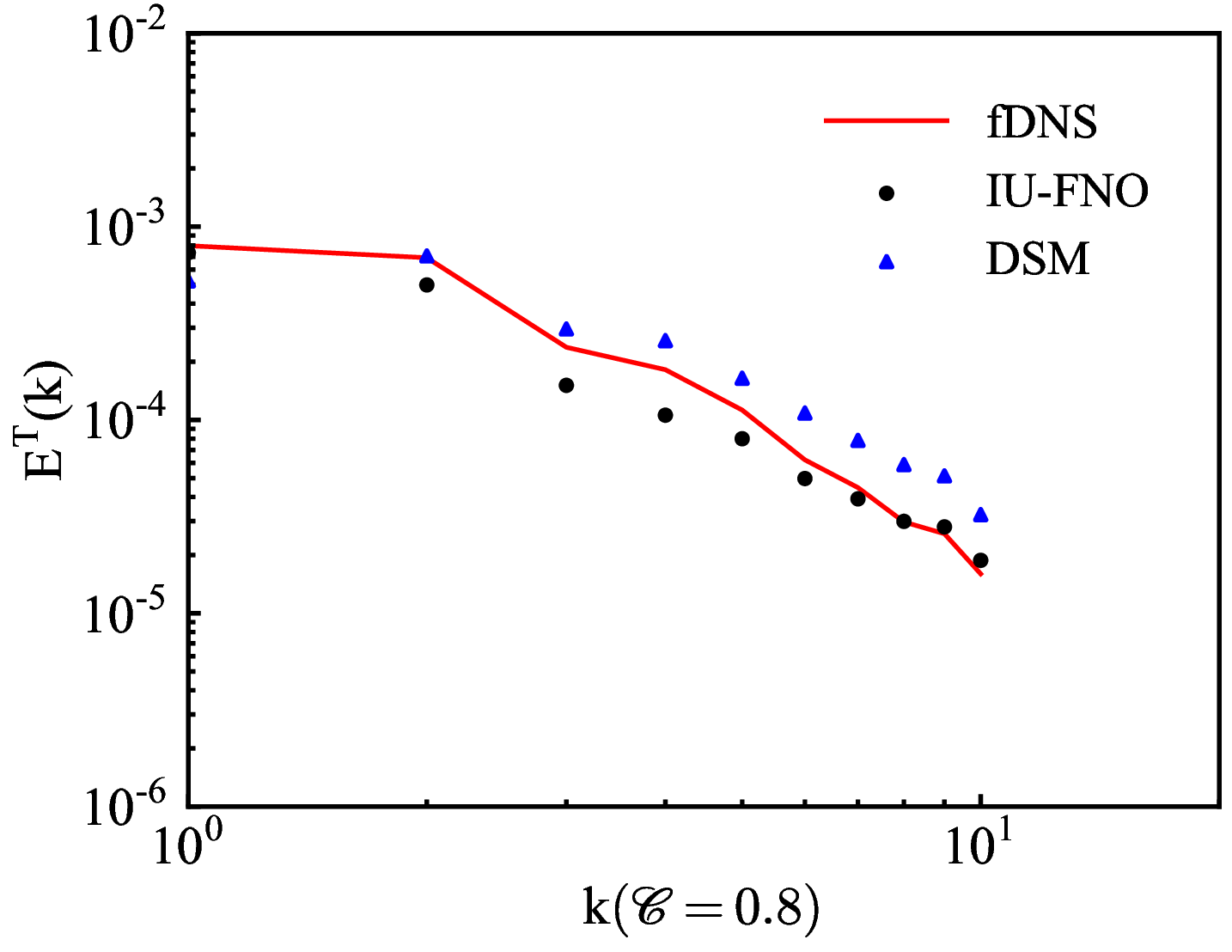}  
    \caption{}  
    \label{spectral4}
  \end{subfigure}
  \begin{subfigure}[b]{0.4\textwidth}
    \centering
    \includegraphics[width=\textwidth]{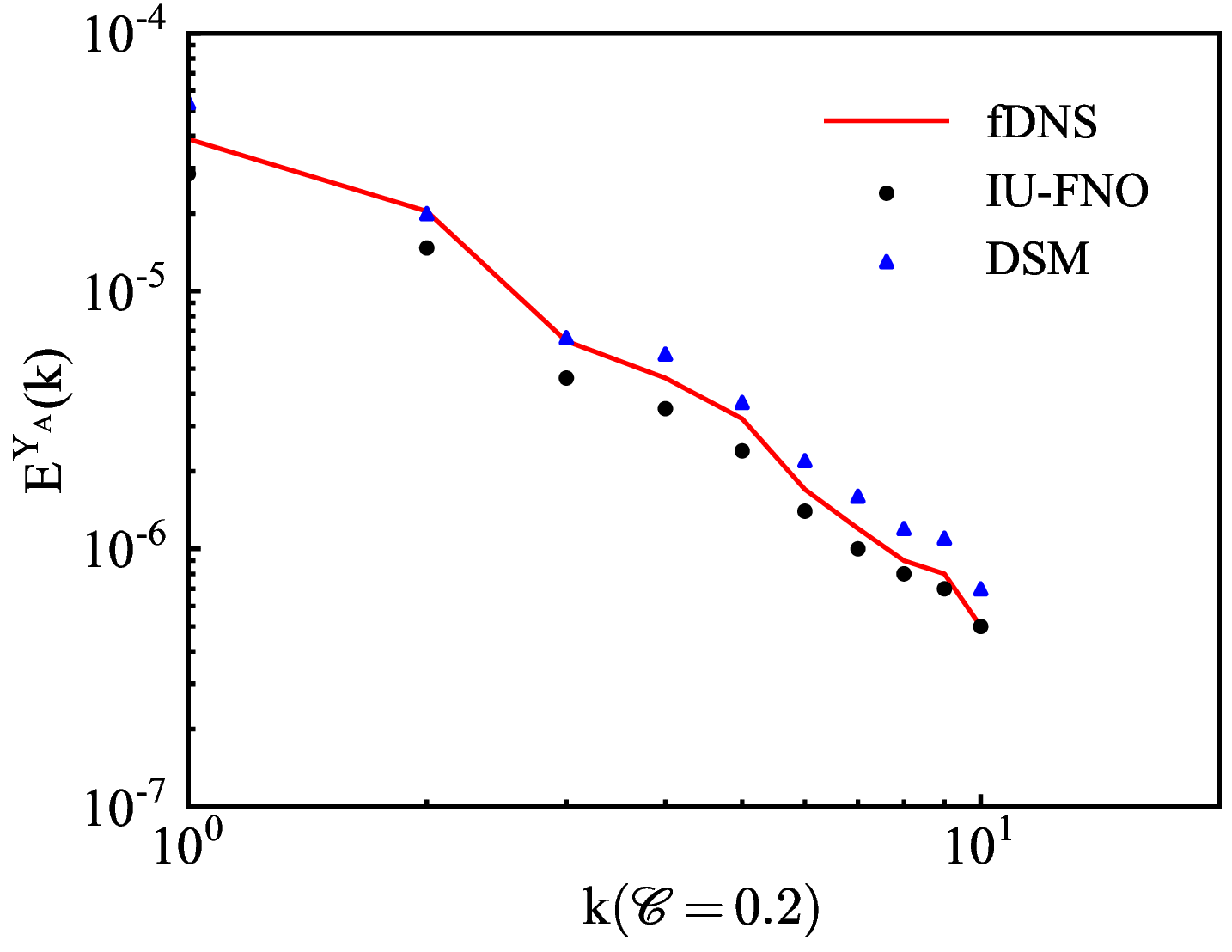}  
    \caption{}  
    \label{spectral5}
  \end{subfigure}
  \begin{subfigure}[b]{0.4\textwidth}
    \centering
    \includegraphics[width=\textwidth]{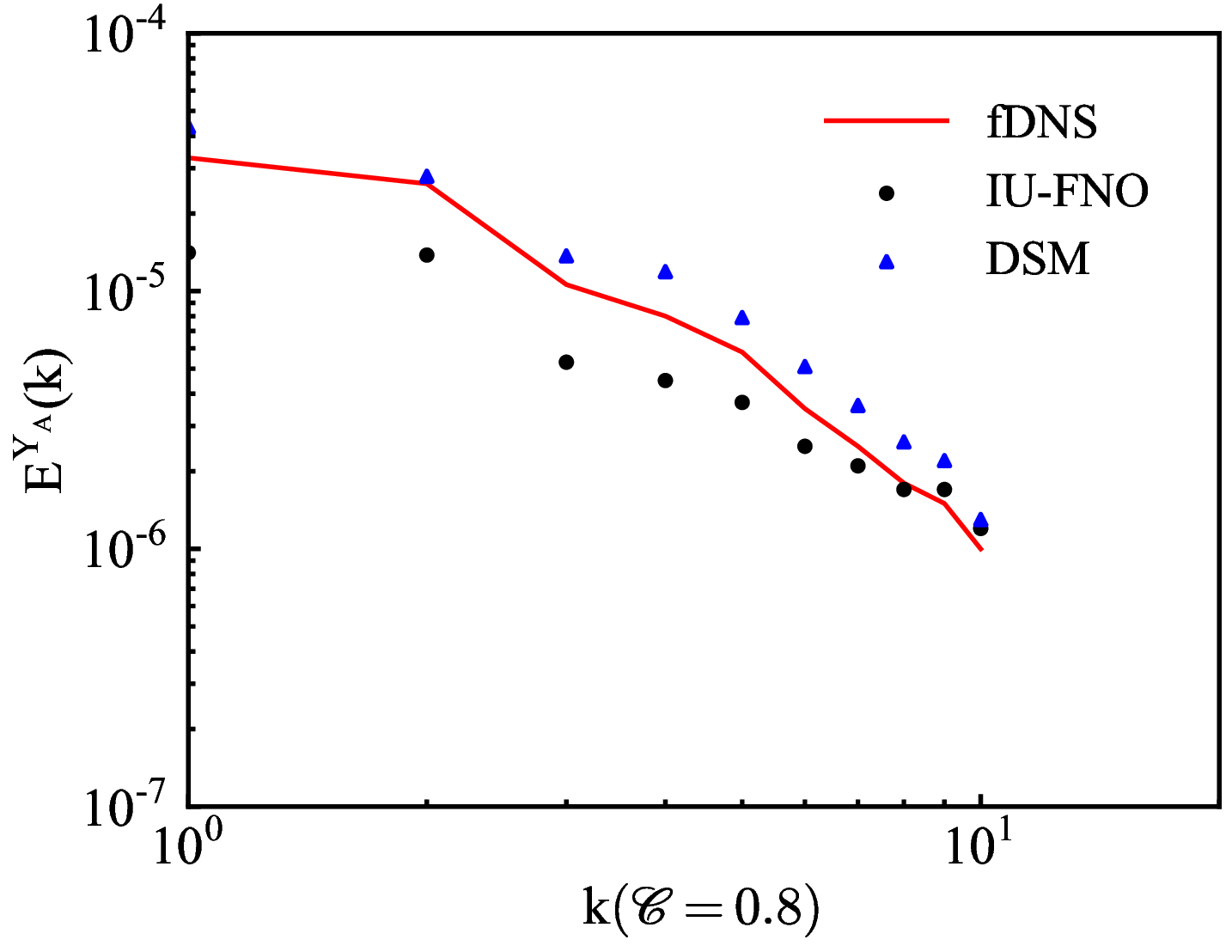}  
    \caption{}  
    \label{spectral6}
  \end{subfigure}
 \caption{Spectra of velocity, temperature and mass fraction of reactant A at different chemically reacting stages in \textit{a posteriori} study.
 (a) the spectrum of velocity at $\mathscr{C} = 0.2$ (b) the spectrum of velocity at $\mathscr{C} = 0.8$ 
 (c) the spectrum of temperature at $\mathscr{C} = 0.2$ (d) the spectrum of temperature at $\mathscr{C} = 0.8$
 (e) the spectrum of mass fraction of reactant A at $\mathscr{C} = 0.2$ (f) the spectrum of mass fraction of reactant A at $\mathscr{C} = 0.8$}
 \label{spectral}
\end{figure}

Fig.~\ref{spectral} illustrates the spectra of velocity, temperature, 
and the mass fraction of reactant for fDNS, IU-FNO and DSM at different 
processes of the reaction. Due to the downsampling in fDNS, only the truncated portion with wave numbers $k\leq10$
is included for comparison. As shown in Figs.~\ref{spectral1} and \ref{spectral2}, the fDNS results for the velocity spectrum generally adhere to the scaling law
$k^{-5/3}$ in the large-scale range. The results indicate that for $k>5$ 
the DSM method exhibits excessive dissipation, while IU-FNO gives a velocity 
spectrum that nearly overlaps with that of fDNS. Furthermore, IU-FNO does not experience the cumulative errors. 
Figs.~\ref{spectral3}-\ref{spectral6} show the spectra 
of temperature and the mass fraction of species A at different times. Overall, the DSM results display 
an overall upward shift in the energy spectrum due to insufficient dissipation in the small 
scales, whereas the IU-FNO results closely align with those of fDNS, maintaining a high precision 
particularly in the small-scale region. However, in Fig.~\ref{spectral6}, the spectrum of mass 
fraction of reactant at $\mathscr{C} = 0.8$ by IU-FNO has larger errors than the DSM when $k \le 4$, 
the cause of such a phenomenon is still unclear.
\begin{figure}[htbp]
  \centering
  \begin{subfigure}[b]{0.49\textwidth}
      \centering
      \includegraphics[width=\textwidth]{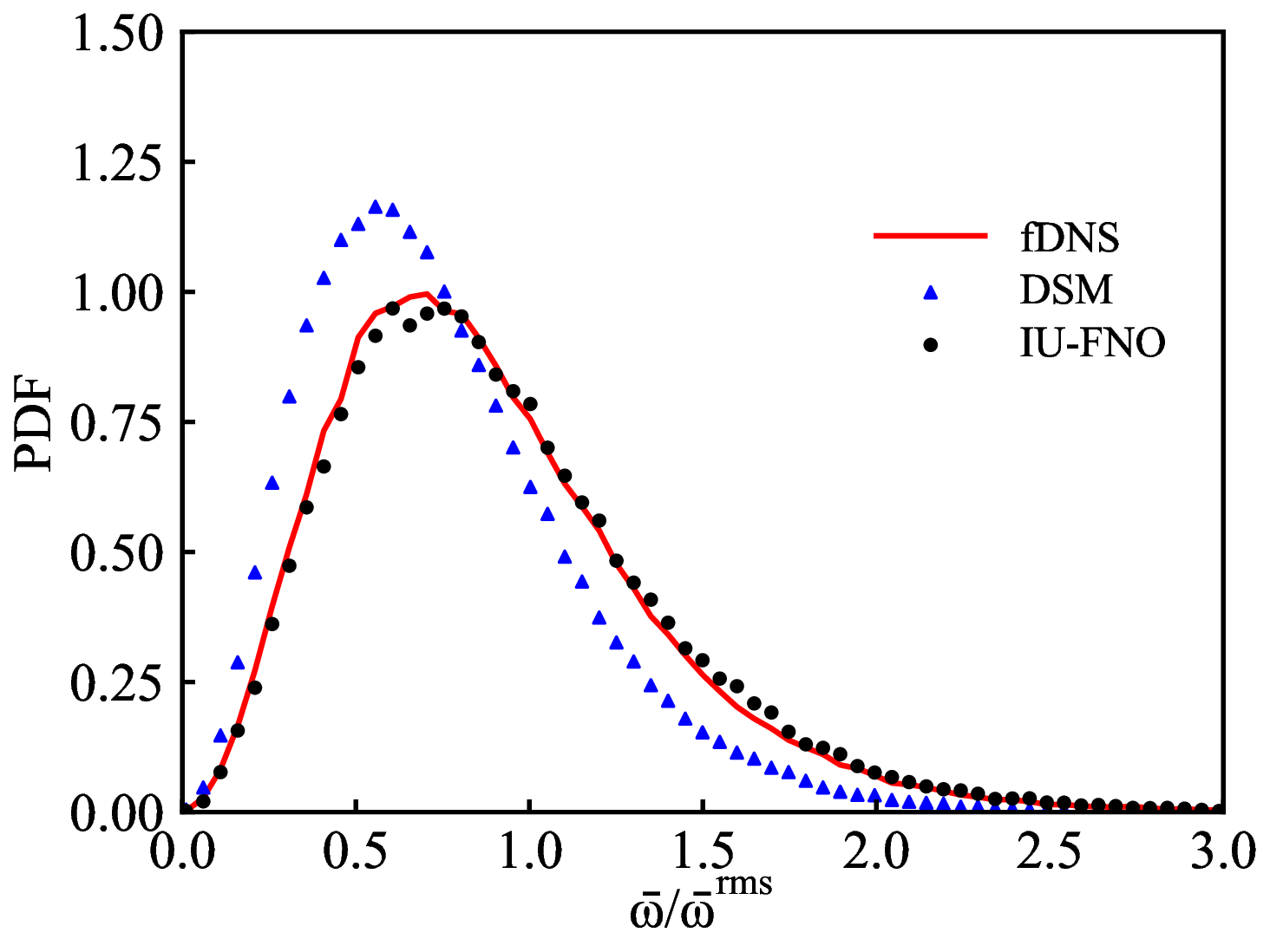}  
      \caption{}  
      \label{vorticity1}
  \end{subfigure}
  \hfill
  \begin{subfigure}[b]{0.49\textwidth}
      \centering
      \includegraphics[width=\textwidth]{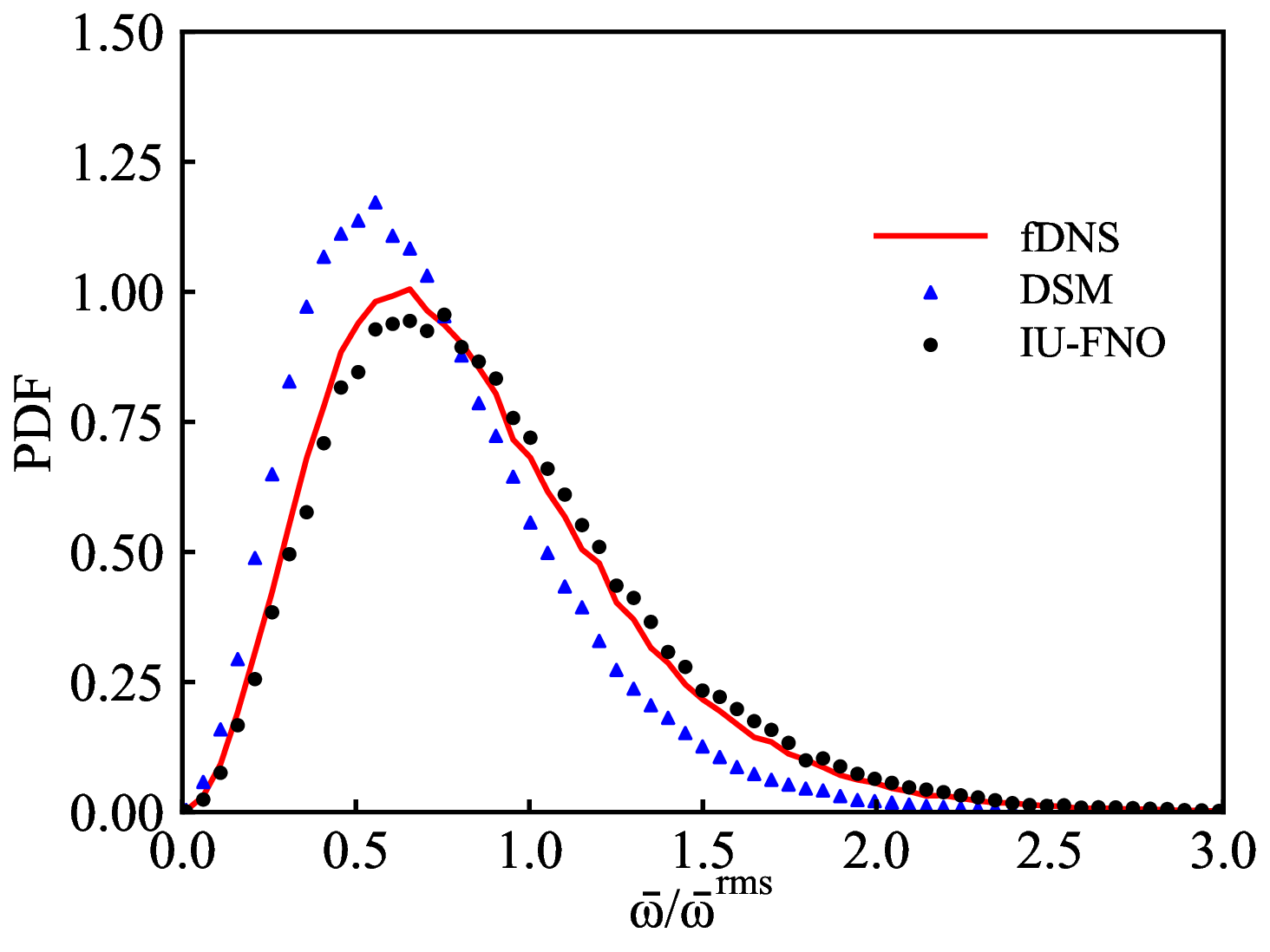}  
      \caption{}  
      \label{vorticity2}
  \end{subfigure}
\caption{PDFs of the normalized vorticity $\bar{\omega}/\bar{\omega}_\mathrm{fDNS}^\mathrm{rms}$ for different models at (a) $\mathscr{C}$ = 0.2 and (b) $\mathscr{C}$ = 0.8 in \textit{a posteriori} study}
\label{vorticity}
\end{figure}

In Fig.~\ref{vorticity}, we present the probability density functions (PDFs) of normalized 
vorticity at different progress of the reaction. 
Here, the vorticity is normalized by the rms values calculated from the fDNS data. The results 
indicate that vorticity statistics predicted by IU-FNO closely match those of fDNS, while the PDFs of DSM 
are higher than fDNS for the range of 0 to 0.8 and lower for normalized vorticity values greater than 0.8. Notably, 
as the reaction progresses, IU-FNO does not exhibit a decrease in accuracy.
\begin{figure}[htbp]
  \centering
  \begin{subfigure}[b]{0.49\textwidth}
      \centering
      \includegraphics[width=\textwidth]{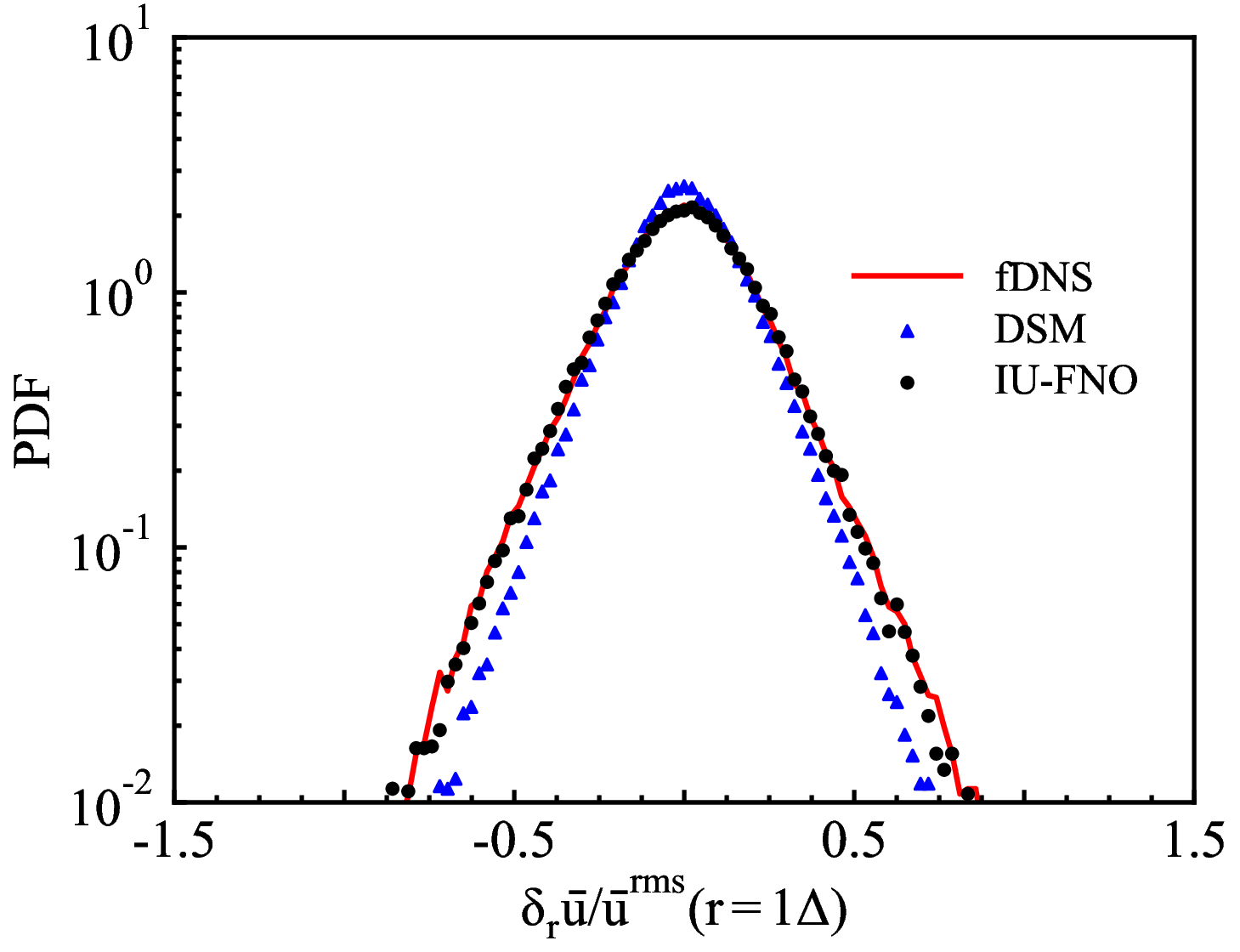}  
      \caption{}  
      \label{inc1}
  \end{subfigure}
  \hfill
  \begin{subfigure}[b]{0.49\textwidth}
      \centering
      \includegraphics[width=\textwidth]{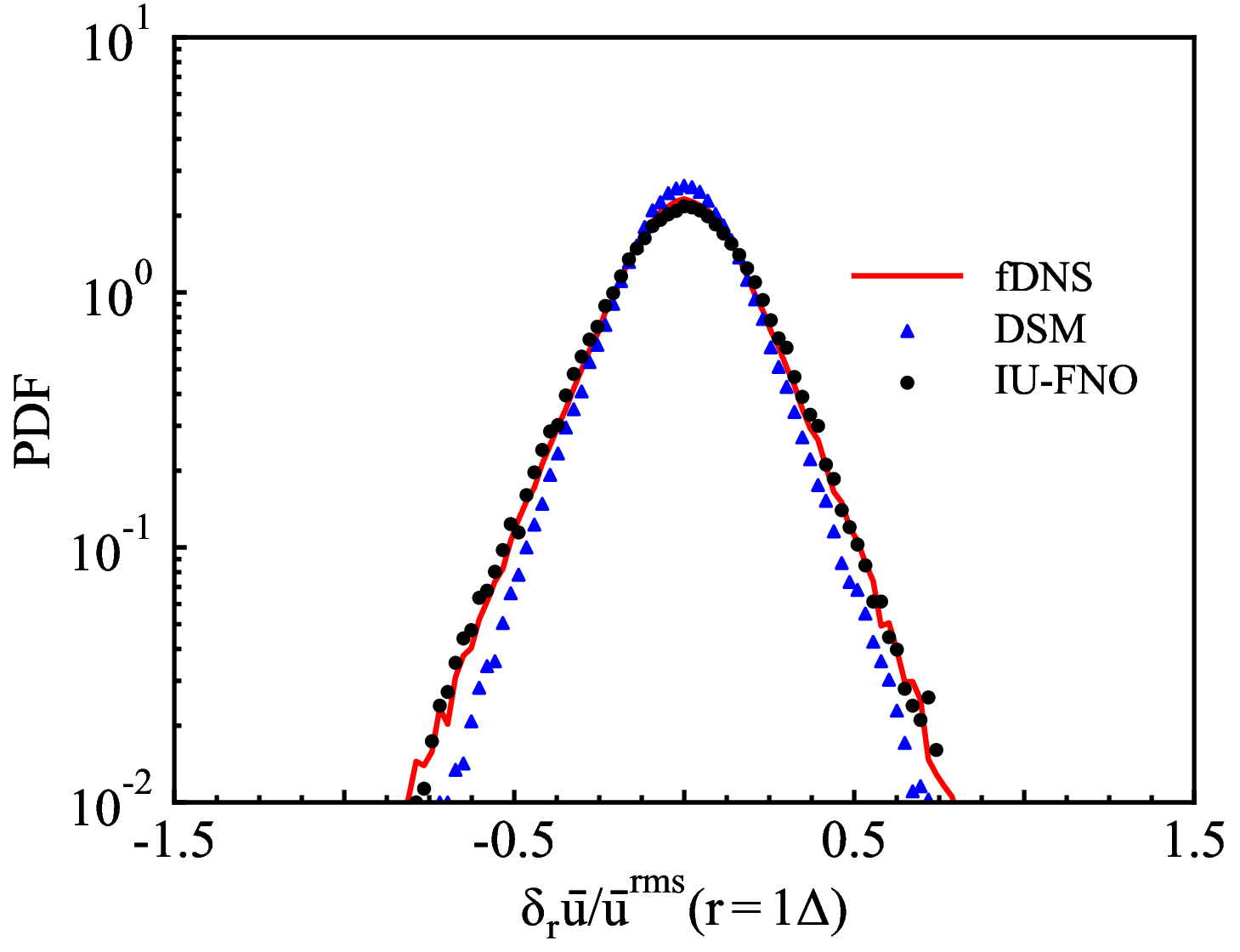}  
      \caption{}  
      \label{inc2}
  \end{subfigure}
\caption{The PDFs of the velocity increment for different models at (a) $\mathscr{C}$ = 0.2 and (b) $\mathscr{C}$ = 0.8 in \textit{a posteriori} study}
\label{inc}
\end{figure}

\begin{figure}[htbp]
  \centering
  \begin{subfigure}[b]{0.49\textwidth}
      \centering
      \includegraphics[width=\textwidth]{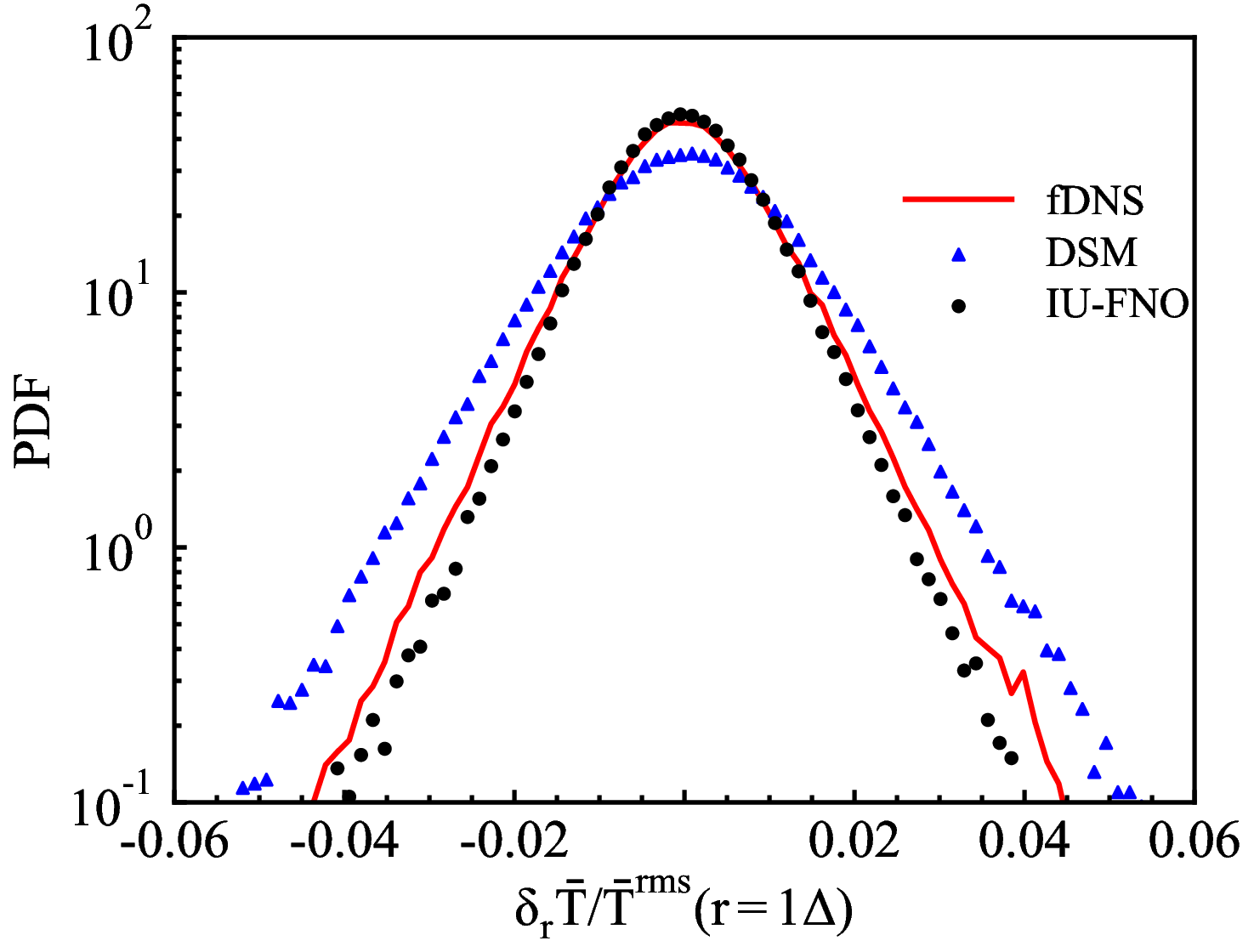}  
      \caption{}  
      \label{incT1}
  \end{subfigure}
  \hfill
  \begin{subfigure}[b]{0.49\textwidth}
      \centering
      \includegraphics[width=\textwidth]{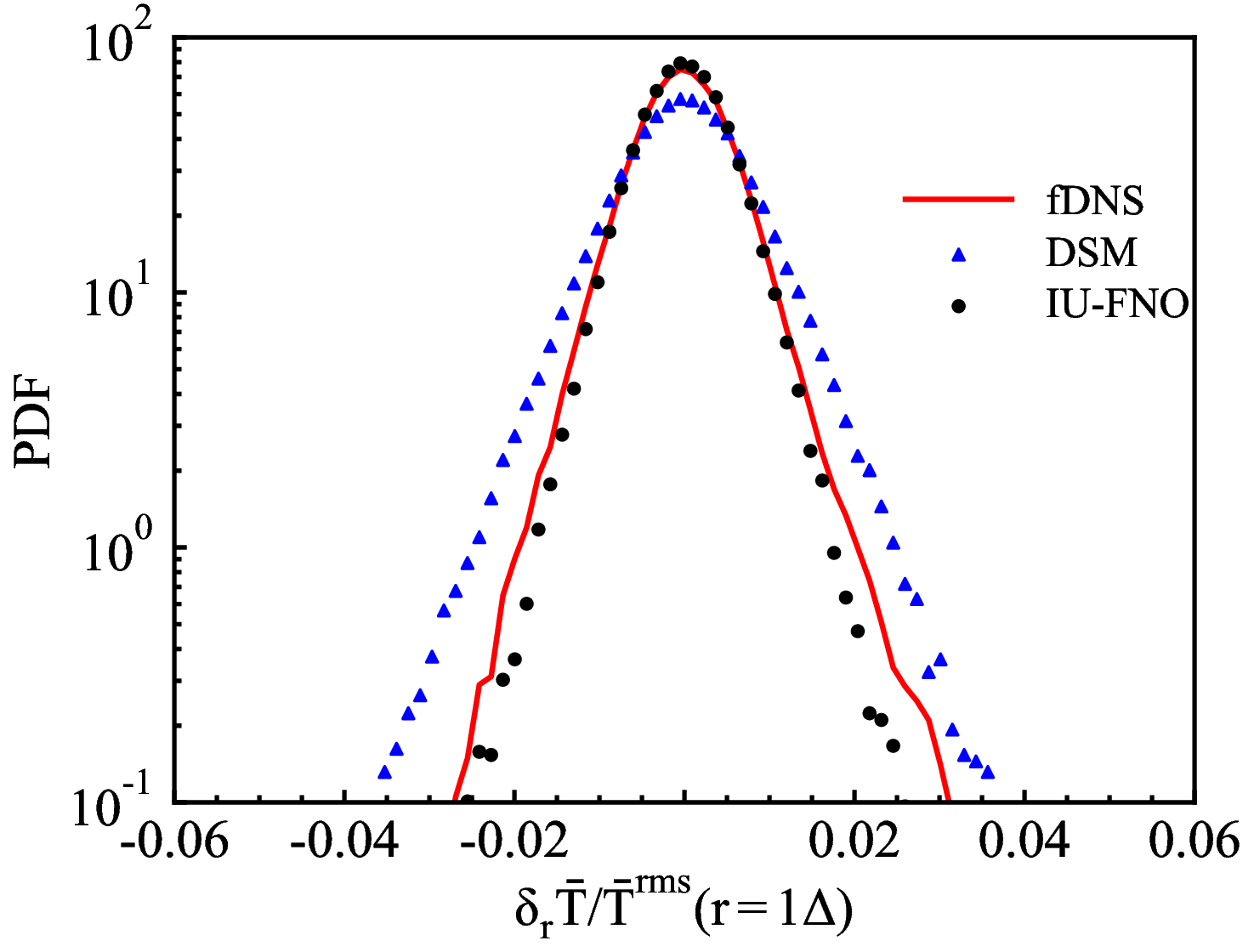}  
      \caption{}  
      \label{incT2}
  \end{subfigure}
\caption{The PDFs of the temperature increment for different models at (a) $\mathscr{C}$ = 0.2 and (b) $\mathscr{C}$ = 0.8 in \textit{a posteriori} study}
\label{incT}
\end{figure}

\begin{figure}[htbp]
  \centering
  \begin{subfigure}[b]{0.49\textwidth}
      \centering
      \includegraphics[width=\textwidth]{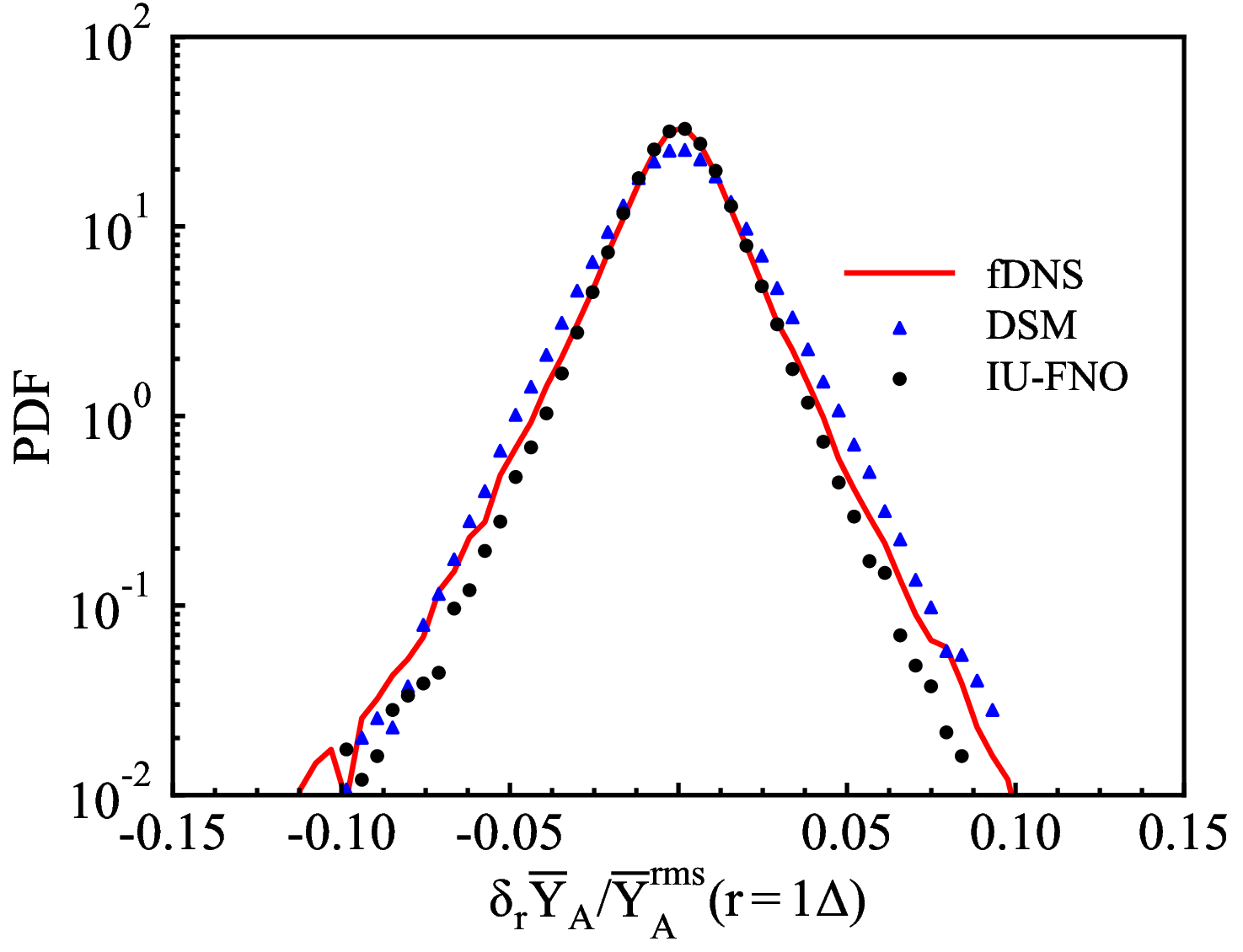}  
      \caption{}  
      \label{incsp1}
  \end{subfigure}
  \hfill
  \begin{subfigure}[b]{0.49\textwidth}
      \centering
      \includegraphics[width=\textwidth]{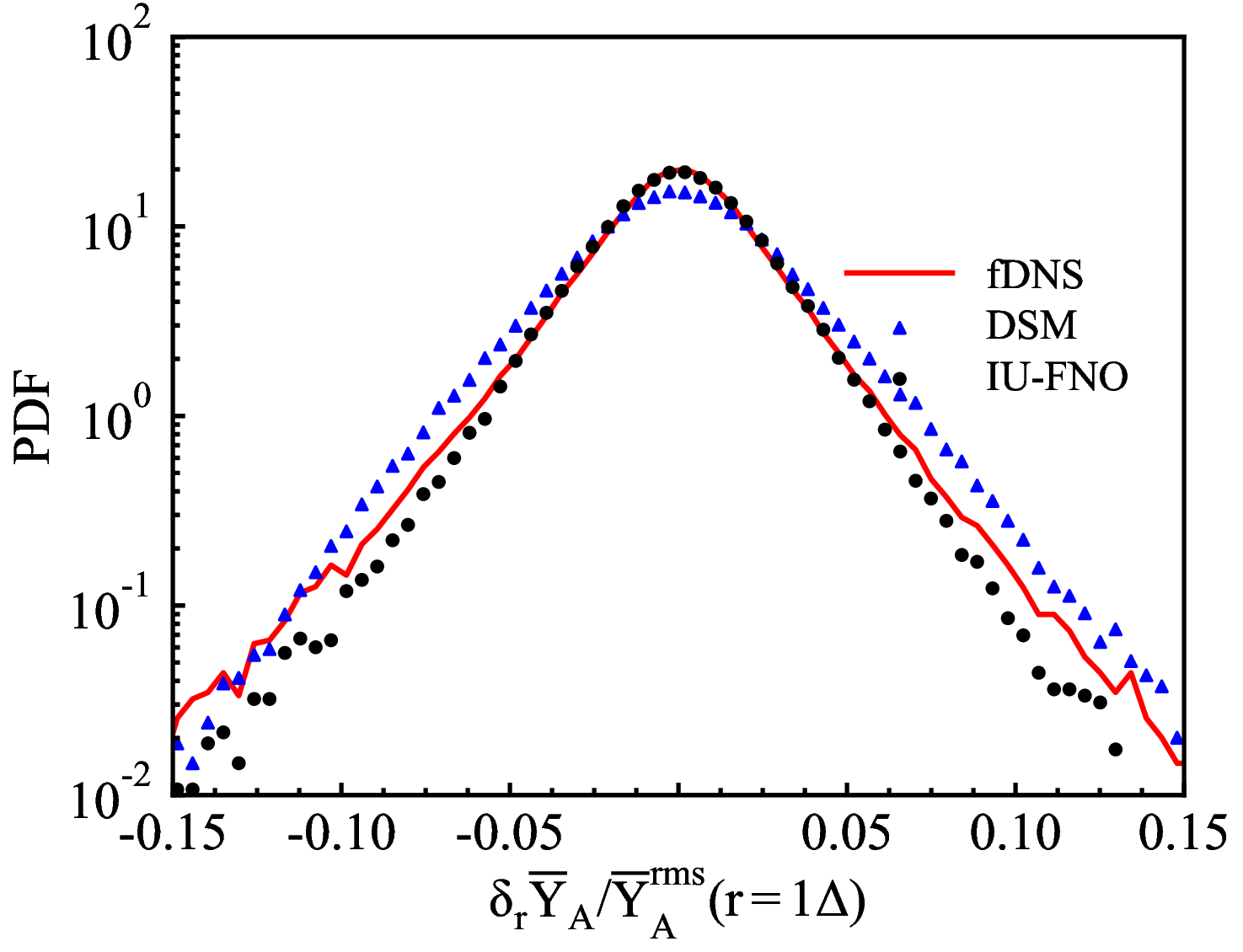}  
      \caption{}  
      \label{incsp2}
  \end{subfigure}
\caption{The PDFs of the mass fraction of reactant A increment for different models at (a) $\mathscr{C}$ = 0.5 and (b) $\mathscr{C}$ = 0.8 in \textit{a posteriori} study}
\label{incsp}
\end{figure}

Furthermore, we compare the probability density functions (PDFs) of the normalized velocity 
increments $\delta_r\bar{u}/\bar{u}^\mathrm{rms}$, temperature increments 
$\delta_r\bar{T}/\bar{T}^\mathrm{rms}$ and mass fraction of reactant A increments 
$\delta_r\bar{Y}_A/\bar{Y}_A ^\mathrm{rms}$ with distance $r = \Delta$ across different stages 
of the chemically reacting in Figs.~\ref{inc}, ~\ref{incT} and ~\ref{incsp}, respectively. Here, 
$\Delta$ denote the distance between two adjacent grid points. 
Figs.~\ref{inc1} and ~\ref{inc2} show that, the PDFs of velocity increments predicted by 
DSM exceed the standard fDNS values near $\delta_r\bar{u}/\bar{u}^\mathrm{rms}$ = 0 
at different stages, while falling below the fDNS values in other regions. 
In contrast, the PDFs of velocity increments predicted by IU-FNO align closely 
with the fDNS values.

From Figs.~\ref{incT} and ~\ref{incsp}, it can be observed that in contrast to the 
PDFs of velocity increments, the PDFs of increments for 
temperature and mass fraction of reactant exhibit significant variation over time. This is attributed to the 
changing intensity of the reaction as time progresses. It is shown that the 
PDFs of the increments predicted by the DSM method deviate from the true values provided by 
fDNS, particularly in the situation of temperature increments, where the discrepancies are 
substantial. In contrast, the IU-FNO predictions align closely with the statistical results from 
fDNS. Only minor divergences occur at the two tails of PDFs.


\begin{figure}[htbp]
  \centering
  \begin{subfigure}[b]{0.4\textwidth}
      \centering
      \includegraphics[width=\textwidth]{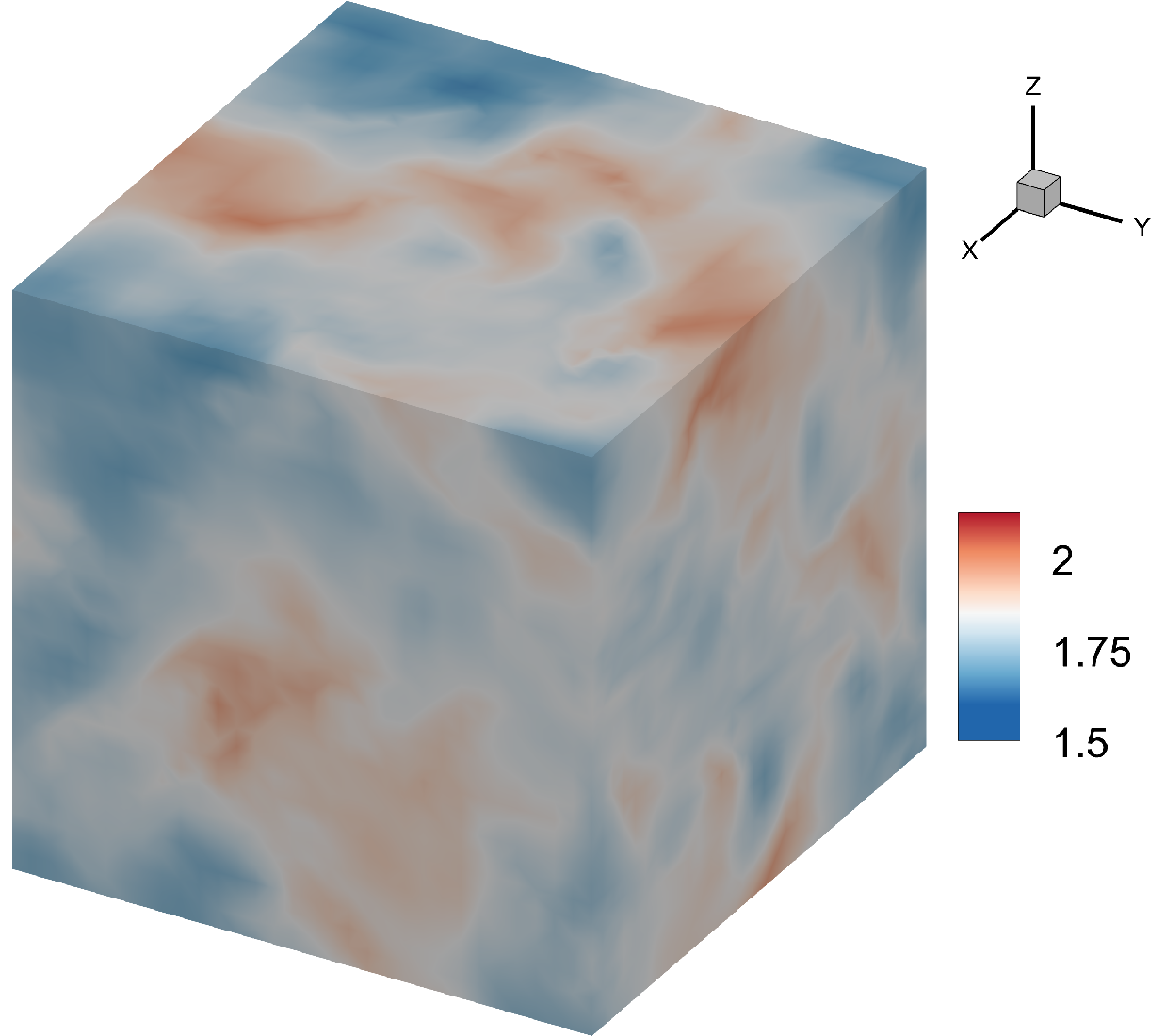}  
      \caption{}  
      \label{contour1}
  \end{subfigure}
  \begin{subfigure}[b]{0.4\textwidth}
      \centering
      \includegraphics[width=\textwidth]{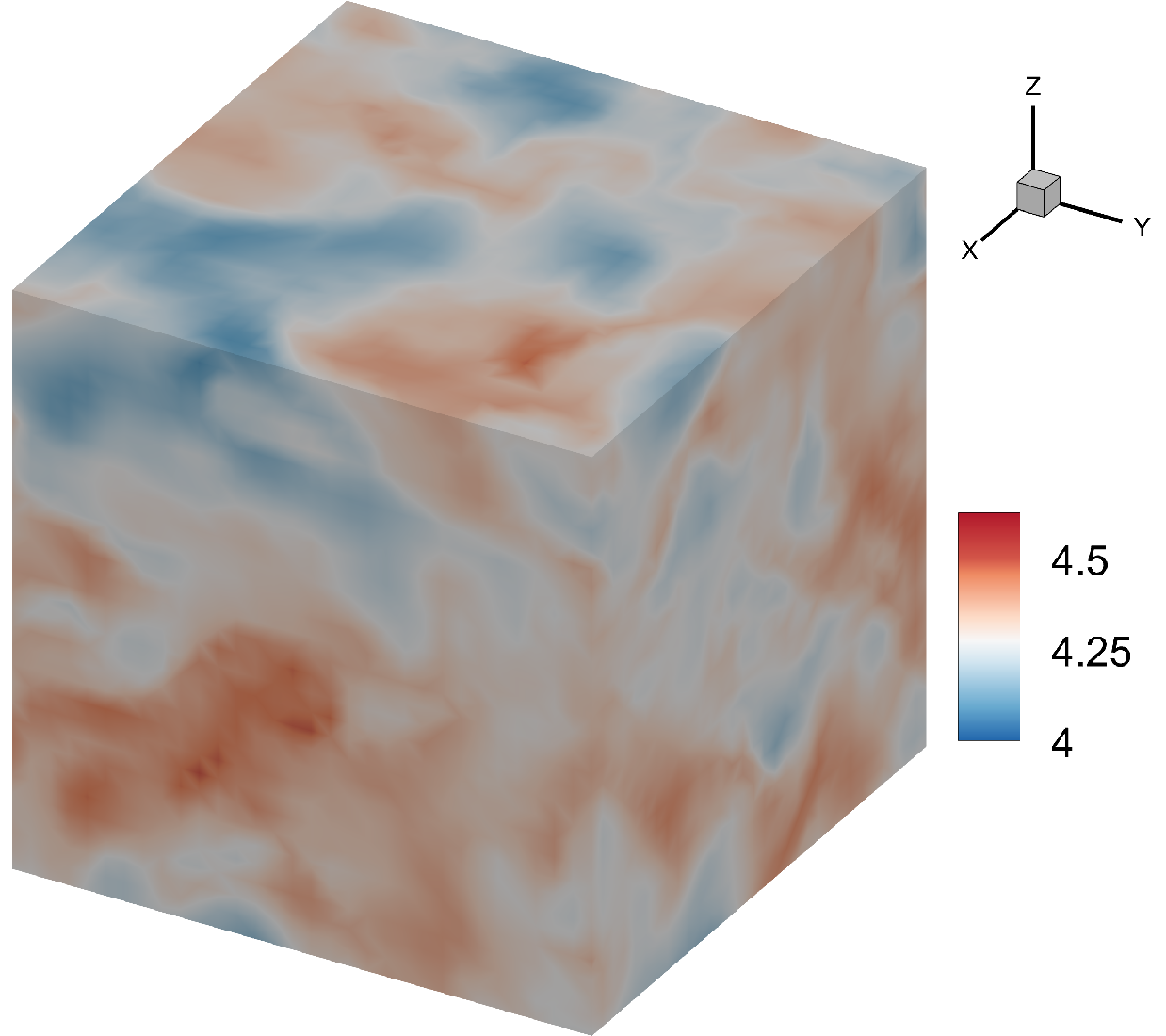}  
      \caption{}  
      \label{contour2}
  \end{subfigure}
  \begin{subfigure}[b]{0.4\textwidth}
    \centering
    \includegraphics[width=\textwidth]{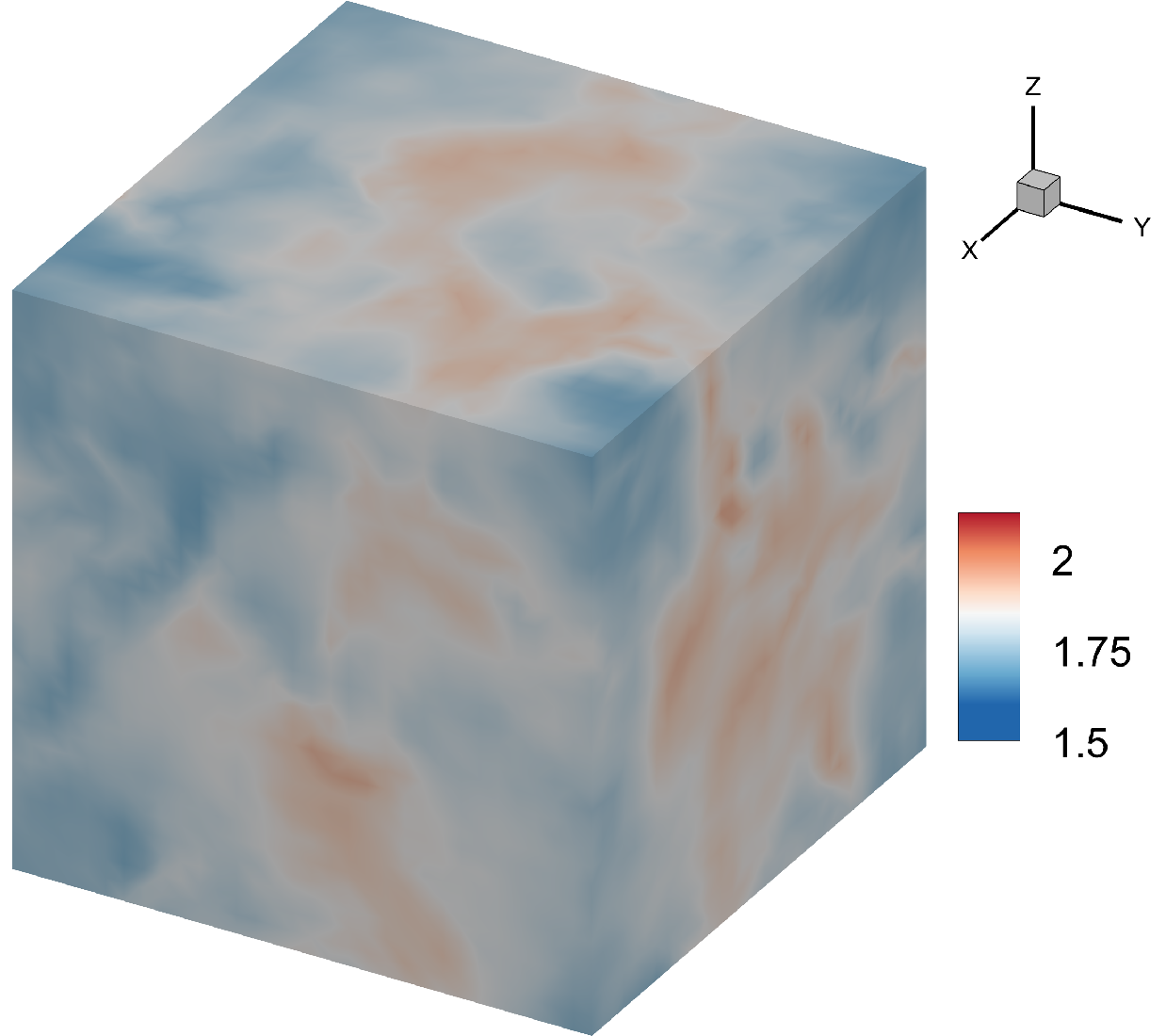}  
    \caption{}  
    \label{contour3}
  \end{subfigure}
  \begin{subfigure}[b]{0.4\textwidth}
    \centering
    \includegraphics[width=\textwidth]{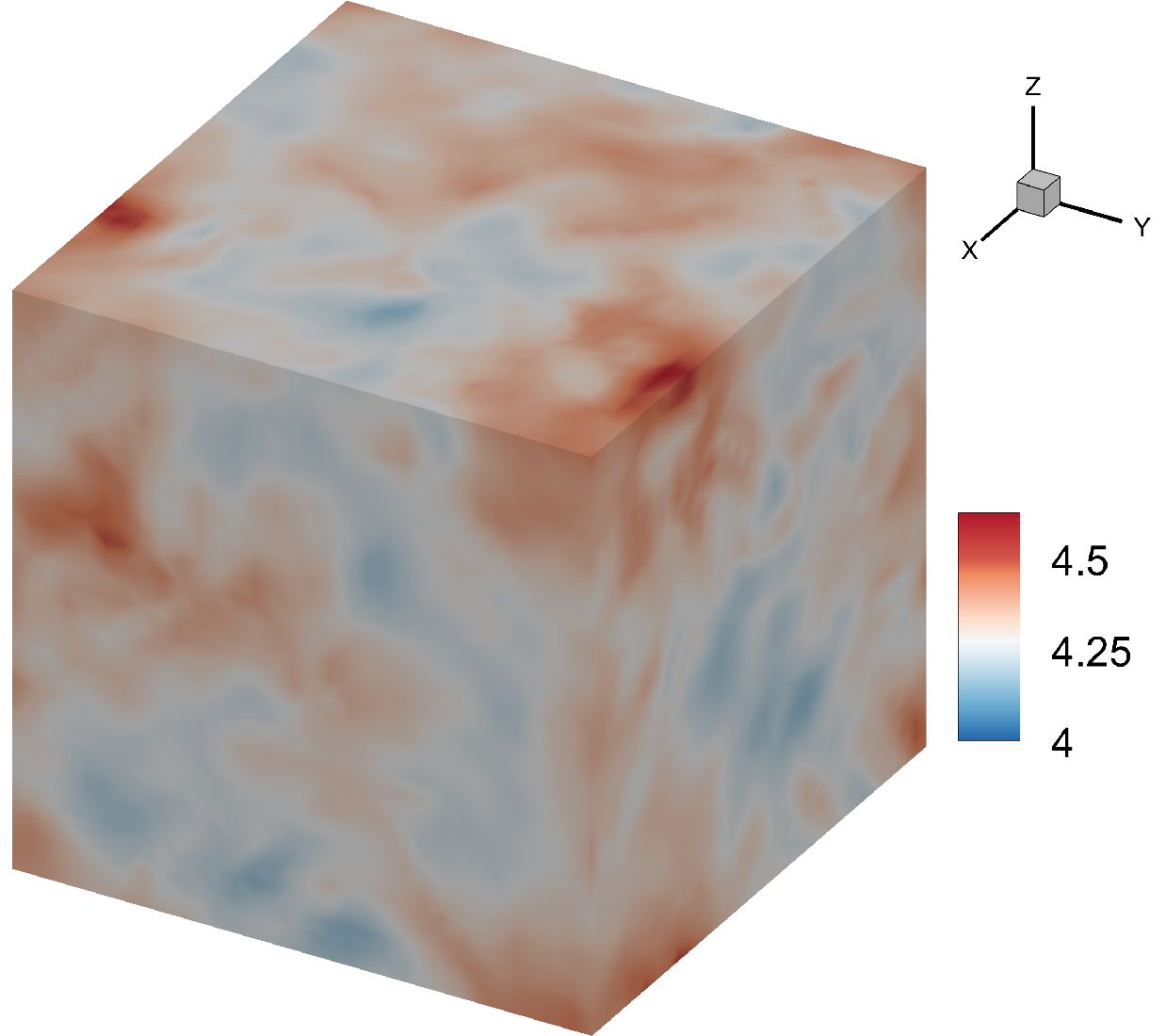}  
    \caption{}  
    \label{contour4}
  \end{subfigure}
  \begin{subfigure}[b]{0.4\textwidth}
    \centering
    \includegraphics[width=\textwidth]{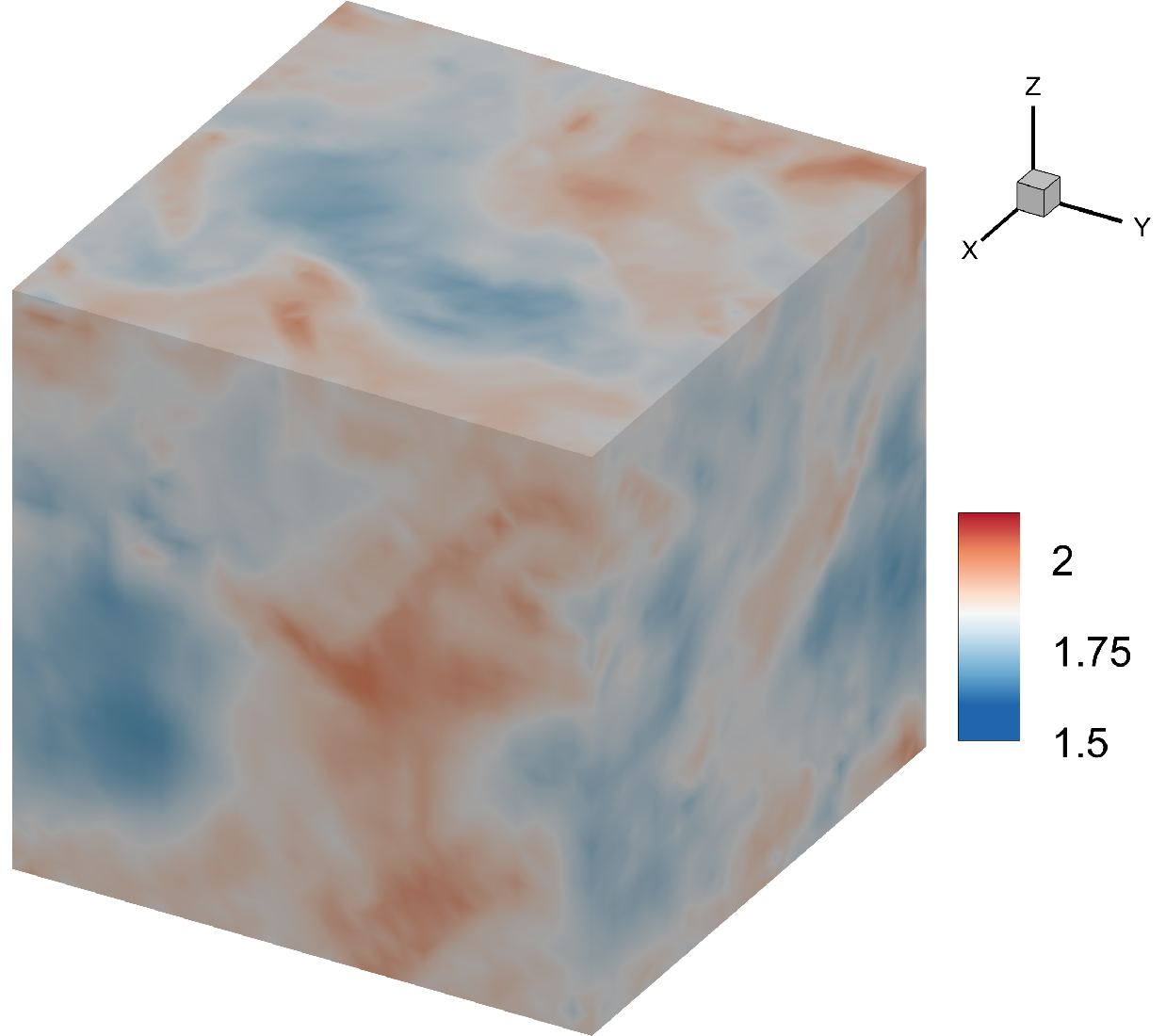}  
    \caption{}  
    \label{contour5}
  \end{subfigure}
  \begin{subfigure}[b]{0.4\textwidth}
    \centering
    \includegraphics[width=\textwidth]{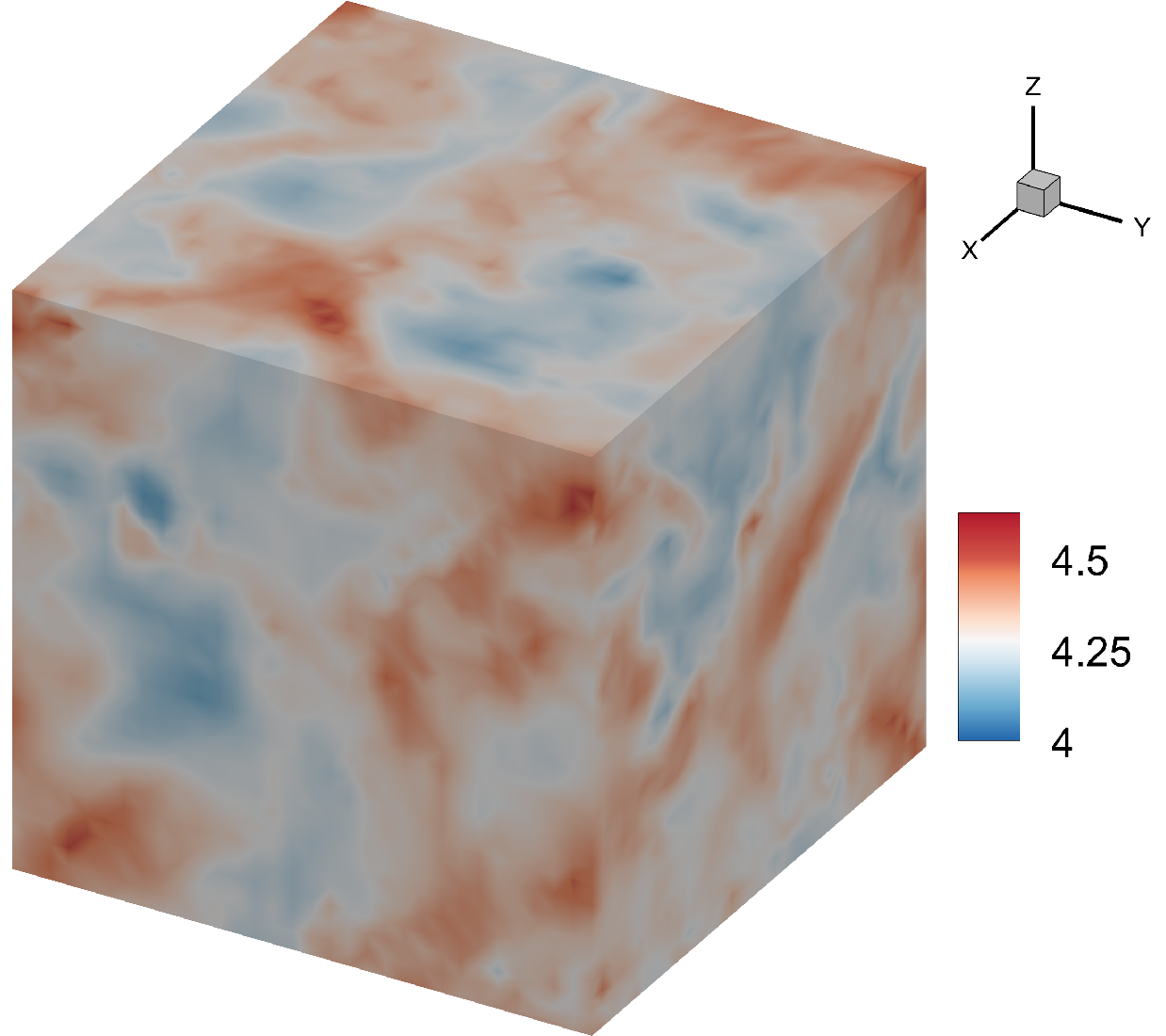}  
    \caption{}  
    \label{contour6}
  \end{subfigure}
 \caption{Contours of temperature T for different models in \textit{a posteriori} study. 
 (a) fDNS at $\mathscr{C}$ = 0.2 and (b) fDNS at $\mathscr{C}$ = 0.8
 (c) IU-FNO at $\mathscr{C}$ = 0.2 and (d) IU-FNO at $\mathscr{C}$ = 0.8
 (e) DSM at $\mathscr{C}$ = 0.2 and (f) DSM at $\mathscr{C}$ = 0.8
 }
  \label{contour}
\end{figure}
Fig.~\ref{contour}  presents the temperature field contours for fDNS, IU-FNO, and DSM at
different time instants. The plots for fDNS, IU-FNO, and DSM are labeled as 
Figs.~\ref{contour1} and~\ref{contour2}, Figs.~\ref{contour3} and~\ref{contour4}, 
Figs.~\ref{contour5} and~\ref{contour6}, respectively. 
The first column shows the contour plots at the reaction progress of $\mathscr{C} = 0.2$, 
while the second column corresponds to $\mathscr{C} = 0.8$. 
It is evident that at different times, fDNS and IU-FNO exhibit similar block distributions. 
In contrast, the DSM results reveal additional flow structures which are not observed in fDNS. 
The turbulent fluctuations in the temperature field predicted by DSM are notably 
stronger compared to fDNS, while IU-FNO’s predictions closely resemble the fluctuations 
observed in fDNS.

  \begin{table}
    \caption{Computational efficiency of different approaches on chemically reacting turbulence.\label{efficiency}}
    \begin{ruledtabular}
    \begin{tabular}{ccc}
    Method & GPU/s & CPU/s \\
    \hline
    DSM & N/A & 157.37 ($\times$ 16 cores) \\
    IU-FNO & 2.14 & 70.06 ($\times$ 1 core) \\
    \end{tabular}
    \end{ruledtabular}
    \end{table}
Table~\ref{efficiency} compares the computational efficiency of the DSM method and the IU-FNO 
method, where the prediction time is the total computational time from the start to the end of 
the reaction. IU-FNO is trained and tested on an Nvidia A100 40G PCIe GPU, while the CPU used 
for loading model parameters and data is an Intel(R) Xeon(R) Gold 6248R CPU @ 3.00 GHz. The DSM simulations are implemented on a computing cluster with Intel Xeon Gold 6148 CPUs, each featuring 16 cores running at 2.40 GHz. 
The Table~\ref{efficiency} illustrates the CPU computation time consumed by different models. 
It is worth noting that IU-FNO 
utilizes a single-core CPU, whereas the DSM method employs a 16-core CPU. Even with 
the parallel computing capability of the 16-core CPU, the total computation time for the DSM 
method exceeds that of the IU-FNO method by more than two times. Thus, IU-FNO demonstrates significant advantages in computational efficiency compared to 
traditional LES methods.

\section{conclusions}\label{6}
In this study, we apply the IU-FNO method to 
chemically reacting compressible turbulence, expanding the number of predicted unknown quantities from the three velocity 
components $u, v, w$ to six variables: the three velocity components $u, v, w$, density $\rho$, temperature $T$, 
and the mass fraction of reactant $Y_A$. These six variables exhibit distinct data 
characteristics within the flow field, increasing the complexity of prediction 
task. To address this challenge, we normalize the data in the preprocessing stage to 
enhance the accuracy of model in predicting all variables.

In terms of forecasting, IU-FNO utilizes a time-stepping iterative approach. 
Specifically, IU-FNO predicts the flow field at the next time step using input data 
that comprises flow field information from the previous five time steps. 
The predicted flow field then replaces the earliest time step in the input, 
forming a new input set for subsequent predictions. 
For network parameter selecting, we identify the optimal parameter set for IU-FNO 
based on training error. 
Regarding training data, we generate 400 cases of chemically reacting turbulence simulations 
from varied initial fields to enhance the capability of generalization for IU-FNO. 
Additionally, five extra cases from varied initial fields are generated for 
\textit{a posteriori} tests.

In the \textit{a posteriori} tests, we compare the results of IU-FNO with the DSM model across 
various metrics, including velocity spectrum, temperature spectrum, mass fraction spectrum of the 
reactant, the probability density functions (PDFs) of vorticity, PDFs of increments of velocity, 
temperature and mass fraction of reactant, and temperature contour plots at different moments. 
It is observed that the results of IU-FNO align closely with those of fDNS, while the DSM 
results exhibit significant deviations. Furthermore, IU-FNO significantly outperforms the 
traditional DSM model in computational efficiency, substantially reducing computational costs.

We also compare the predictive performance of FNO and IU-FNO. Due to the significant cumulative 
in FNO during time advancement, combined with the inherent coupling of flow and 
temperature variations in three-dimensional compressible reactive turbulent flows, 
FNO tends to diverge soon after a few prediction steps. Consequently, it fails to 
adequately predict 
the entire chemically reacting process and hence we have not presented the 
predictive results 
from FNO.

For future work, we have several prospects for the model. We aim to incorporate 
constraints from the Navier-Stokes equations to reduce dependence of model on data and 
enhance the 
generalization performance of model across different flow fields. Additionally, we have only tested 
the chemically reacting turbulence at relatively low Mach numbers; there remains significant 
potential for neural operators in predicting turbulent flow fields at higher Mach numbers.

\section{data availability}\label{7}
The data that support the findings of this study are available from the corresponding 
author upon reasonable request.

\begin{acknowledgments}
  This work was supported by the National Natural Science	Foundation of China (NSFC Grant Nos. 12172161, 12302283, 92052301, and 12161141017), by the Shenzhen Science and Technology Program (Grant No. KQTD20180411143441009), and by Department of Science and Technology of Guangdong Province (Grant No. 2019B21203001, No. 2020B1212030001, and No. 2023B1212060001). This work was also supported by Center for Computational Science and Engineering of Southern University of Science and Technology, and by National
  Center for Applied Mathematics Shenzhen (NCAMS).
\end{acknowledgments}

\nocite{*}
\bibliography{aipsamp}

\end{document}